\documentclass[a4paper,11pt]{article}
\pdfoutput=1
\usepackage{jheppub}
\usepackage[utf8]{inputenc}
\usepackage[english]{babel}
\usepackage{amsmath,amssymb}
\usepackage{mathrsfs}
\usepackage{mathpazo}
\usepackage{float}

\title{\boldmath The SU(3) Sextet Model with Wilson Fermions}

\author[a]{Martin Hansen}
\author[b]{Vincent Drach}
\author[a]{Claudio Pica}
\affiliation[a]{CP$^3$-Origins, University of Southern Denmark, Campusvej 55, DK-5230 Odense M, Denmark.}
\affiliation[b]{Centre for Mathematical Sciences, Plymouth University, Plymouth, PL4 8AA, United Kingdom.}
\emailAdd{hansen@cp3.sdu.dk}
\emailAdd{vincent.drach@plymouth.ac.uk}
\emailAdd{pica@cp3.sdu.dk}

\makeatletter
\def\@fpheader{\relax~}
\makeatother

\abstract{
  We investigate the spectrum and IR properties of the SU(3) ``sextet" model with two Dirac fermions in the two-index symmetric representation via lattice simulations. 
  This model is a prime candidate for a realization of Walking Technicolor, which features a minimal matter content and it is expected to be inside or very close to the lower boundary of the conformal window.   
  We use the Wilson discretization for the fermions and map the phase structure of the lattice model. 
  We study several spectral and gradient flow observables both in the bulk and the weak coupling phases.
  While in the bulk phase we find clear signs of chiral symmetry breaking, in the weak coupling phase there is no clear indication for it, and instead the chiral limit of the model seems compatible with an IR-conformal behavior.
 }

\preprint{CP3-Origins-2017-019 DNRF90}

\newcommand{\re}{\mathrm{Re}}
\newcommand{\tr}{\mathrm{Tr}}
\newcommand{\bra}[1]{\langle #1\vert}

\begin{document}
\maketitle
\flushbottom

\section{Introduction}
\label{sec:intro}
The discovery of a new scalar state in 2012 was a tremendous success for the Large Hadron Collider and for the Standard Model. The properties of this new scalar state have since been studied by the ATLAS and CMS collaborations and they are in general agreement with the SM prediction for an elementary Higgs boson.

Although it is an economical description, from a theoretical standpoint the Higgs sector is unappealing.
In fact, the Higgs sector is by many regarded as a model of spontaneous EW symmetry breaking, rather than a true dynamical explanation. Moreover, the electroweak scale is not protected against quantum corrections, which makes the model unnaturally fine-tuned.

The discovery of the Higgs boson, with properties closely resembling the SM elementary Higgs, excludes a large number of BSM models, such as the ``Higgsless models'', and the traditional Technicolor theories, based on QCD-like dynamics. However, a wide class of composite Higgs theories, in which EW symmetry is broken dynamically by a new strong force, are still compatible with the experiments.

The two most interesting realizations of composite Higgs models are Walking Technicolor (WTC)~\cite{Weinberg:1979bn,Susskind:1978ms,Dimopoulos:1979es,Eichten:1979ah,Farhi:1980xs,Holdom:1984sk,Yamawaki:1985zg,Bando:1986bg,Appelquist:1987fc,Caswell:1974gg,Banks:1981nn,Sannino:2004qp,Dietrich:2006cm} and pNGB Higgs models~\cite{Kaplan:1983fs,Kaplan:1983sm,Banks:1984gj,Georgi:1984ef,Georgi:1984af,Dugan:1984hq}. 
In such models the Higgs is regarded as the pseudo Nambu-Goldstone boson (pNGB) of an approximate global symmetry, which explains the little hierarchy between the mass of the Higgs and the other resonances of the strong sector. This extra symmetry is a global flavor symmetry in the case of pNGB Higgs models, or an approximate scale invariance symmetry in the case of Walking Technicolor.

Walking Technicolor models are asymptotically free models which can be considered as a small deformation of a conformal field theory in the infrared. Such models share several important features, which makes them good candidates for a composite Higgs model. This includes the possible emergence of a light $0^{++}$ scalar state, associated to the approximate scale invariance, that can play the role of the Higgs boson, with a light mass and couplings similar to the SM Higgs. The strong coupling might evolve slowly with the energy scale (i.e. it \textit{walks}), and if the model also has a large mass anomalous dimension $\gamma\sim1-2$, SM fermion masses could be generated without large Flavor Changing Neutral Currents. Furthermore, models with only two EW gauged fermions are favored as they have a smaller $S$-parameter and do not violate constraints on EW precision tests \cite{Sannino:2004qp}.

Another reason to consider WTC models is that any four-dimensional fundamental realization of the composite pNGB Higgs admits a Technicolor-like limit and fine-tuning is needed to align the vacuum in the Goldstone Higgs direction \cite{Cacciapaglia:2014uja}. 

Here we study the so-called ``sextet model", a WTC model based on an SU(3) gauge theory with a doublet of Dirac fermions in the two-index symmetric (sextet) representation. From perturbation theory this model is expected to be inside the conformal window~\cite{Ryttov:2017kmx}, although the large anomalous dimension could trigger chiral symmetry breaking, pushing the model outside~\cite{Sannino:2004qp}. 

Non-perturbative calculations are required to settle this issue and confirm or exclude this model as being phenomenologically viable. If the model is chirally broken, a non-perturbative determination of the mass of the lightest scalar state and the spin-1 resonances will provide crucial input for searches at the LHC.  If the theory is conformal, the existence of four-fermion interactions can drive the theory away from conformality \cite{Fukano:2010yv,Rantaharju:2016jxy,Rantaharju:2017eej}. 

The sextet model has been studied previously by several groups \cite{
  Shamir:2008pb,DeGrand:2008kx,DeGrand:2010na,Kogut:2010cz,Kogut:2011ty,DeGrand:2012yq,Kogut:2015zta,Hasenfratz:2015ssa,
  Fodor:2012ty, Fodor:2014pqa, Fodor:2016wal, Fodor:2016pls}.
The non-perturbative $\beta$-function for the model has been calculated, in similar schemes, both using staggered fermions and improved Wilson fermions. 
The results for the two kind of discretizations are in tension, with the Wilson fermion simulations pointing to the existence of an IR fixed point. Results for the non-perturbative $\beta$-function in (near) IR conformal model should, however, be taken with extreme care. 
In fact, close to an (approximate) IR fixed point very large volumes are needed for a reliable continuum extrapolation, as pointed out in \cite{Fodor:2015qzl,Fodor:2016zil}.

Recent studies of the spectrum with improved rooted staggered fermions have a preference towards chiral symmetry breaking. 
Several states in the spectrum were studied, including the Goldstone sector, the vector and axial vector mesons, the nucleon and both the isosinglet and isotriplet scalars.

A striking property of the spectrum is that, while hadron masses and decay constants depend strongly on the quark mass, ratios like $m_X/f_{PS}$, where $m_X$ is the mass of a hadron and $f_{PS}$ is the pseudoscalar decay constant, appear to be approximately independent of the quark mass. 
This would indicate an IR conformal behavior.
However it was shown that the numerical data for the spectrum is not well described by the leading order scaling behavior expected in the IR-conformal hypothesis. 

On the other hand, it seems possible to fit the observed Goldstone spectrum by using rooted staggered chiral perturbation theory and the spectral density of the Dirac operator shows a plateau at small eigenvalues. However it is unclear if the value of the chiral condensate from the GMOR relation and the one from the Banks-Casher relation are compatible. 

A light $0^{++}$ scalar state is present in the spectrum, which is in fact lighter than the would-be Goldstone boson over the entire range of light quark masses explored. This casts doubts about the applicability of chiral perturbation theory to model the numerical data.

Not many studies for the spectrum of the model are available which use the Wilson discretization. 
Given the tension between studies with staggered and Wilson fermions, we started a more thorough investigation of the spectrum with Wilson fermions \cite{Drach:2015sua,Hansen:2016sxp}.
Here we present our final results for the spectrum. In Section~\ref{sec:lattice} we describe our lattice setup for the numerical computation, in Section~\ref{sect:phases} we report our findings for the phase structure of the lattice model, and finally in Section~\ref{sect:wcp} we discuss the spectrum of the model. 

\section{Lattice formulation}
\label{sec:lattice}
The non-perturbative simulations are performed after introducing a UV and IR cutoff in the form of a space-time lattice of finite extent $L^3\times T$. The Euclidean formulation of the theory is used and the path integral is thus reduced to an ordinary integral over a large number of degrees of freedom. In this section we outline our simulation strategy, the simulated action, and the calculated observables.

\subsection{Action}
\label{sec:action}
The choice of a discretized action on the lattice is not unique, nor is the use of boundary conditions. For the gauge field we use the standard plaquette action \cite{Wilson:1974sk}
\begin{equation}
 S_G[U] = -\frac{\beta}{N_c}\sum_x\sum_{\mu<\nu}\re~\tr~P_{\mu\nu}(x),
\end{equation}
where $N_c$ is the number of colors and $\beta=2N_c/g^2$ is the inverse bare coupling. The elementary plaquette $P_{\mu\nu}(x)$ in the ($\mu$,$\nu$)-plane at position $x$ is defined as
\begin{align}
P_{\mu\nu}(x) &= U_\mu(x)U_\nu(x+\hat{\mu})U_\mu^\dagger(x+\hat{\nu})U_\nu^\dagger(x).
\end{align}
We impose periodic boundary conditions on the gauge field in all directions. For the fermions we use the the unimproved Wilson-Dirac operator \cite{Wilson:1974sk}
\begin{equation}
 D_W[U^R] = m_0 + \frac{1}{2}\sum_\mu\{\gamma_\mu(\nabla_\mu+\nabla^*_\mu) -a\nabla_\mu^*\nabla_\mu\},
\end{equation}
where $m_0$ is the bare quark mass and $a$ is the lattice spacing. The operators $\nabla_\mu$ and $\nabla_\mu^*$ are the discretized forward and backward covariant derivatives, respectively.
\begin{align}
 \nabla_\mu\psi(x) &= U_\mu^R(x)\psi(x+\hat{\mu}) - \psi(x) \\
 \nabla_\mu^*\psi(x) &= \psi(x) - U_\mu^R(x-\hat{\mu})^\dagger\psi(x-\hat{\mu})
\end{align}
In these definitions $U_\mu^R(x)$ is the parallel transporter (link) from a site $x$ to its neighbor at $x+\hat{\mu}$ in the representation $R$. In the case of two mass degenerate flavors, the fermion action can be written as
\begin{equation}
 S_F[U^R,\bar{\psi},\psi] = a^4\sum_{f=1}^2\sum_x\bar{\psi}_f(x)D_W[U^R]\psi_f.
\end{equation}
For the fermions we impose periodic boundary conditions in the spatial directions and anti-periodic boundary conditions in the temporal direction.

With this choice of discretization, the action only depends on two bare parameters: the inverse gauge coupling $\beta$ and the (dimensionless) quark mass $am_0$. While the link variables $U_\mu(x)$ appearing in the gauge action are in the fundamental representation of the gauge group, the links in the Wilson-Dirac operator $U_\mu^R(x)$ are in the same representation $R$ as the fermion fields.

For the two-index symmetric representation, the mapping between the fundamental and the represented links is given by
\begin{equation}
 (U^R)^{ab} = \tr[e^aUe^bU^T],
\end{equation}
where $\{e^a\}$ is the orthonormal basis for the representation (see Appendix~\ref{sect:sextet}).

\subsection{Algorithm}
\label{sec:algoritm}
The results presented in this work are obtained using our own simulation code first described in \cite{DelDebbio:2008zf}. The algorithm used is the standard Hybrid Monte Carlo (HMC) algorithm \cite{Duane:1987de} together with various improvements, such as even-odd preconditioning \cite{DeGrand:1990dk}, mass preconditioning \cite{Hasenbusch:2001ne}, second order OMF integrator \cite{OMF}, and chronological inversion \cite{Brower:1995vx}.

For some of the simulations we use a GPU accelerated version of the code, however, because the GPU implementation lacks support for parallelism its use has been limited to smaller volumes and/or heavy masses.

\subsection{Observables}
\label{sec:observables}
We report in this section the definition of the observables used when studying the model. The simplest observable measured is the average value of the plaquette $P_{\mu\nu}(x)$ over all space-time points and possible orientations
\begin{align}
\langle P \rangle &= \frac1{6 N_c V}\sum_x \sum_{\nu<\mu}\re~\tr~P_{\mu\nu}(x)\, .
\end{align}

\subsubsection*{Mesons}
To extract the masses of the isotriplet mesons, their decay constants, and the quark mass as defined
from the Partially Conserved Axial Current (PCAC) relation, we consider two-point functions at zero momentum.
\begin{equation}
 C_{\Gamma\Gamma'}(t)
 = \sum_\mathbf{x} \langle(O_\Gamma^\dagger(\mathbf{x},t)O_{\Gamma'}(0)\rangle
\end{equation}
The interpolating operator is given by $O_\Gamma(\mathbf{x},t)=\bar{\psi}_1(\mathbf{x},t)\Gamma
\psi_2(\mathbf{x},t)$ where $\psi_1$ and $\psi_2$ represent two different flavors of mass-degenerate fermion fields, and $\Gamma$ is a generic matrix in the Clifford algebra. For the hadronic quantities used in this work, $\Gamma$ and $\Gamma'$ range over the choices $\{\gamma_5,\gamma_0\gamma_5,\gamma_k,\gamma_5\gamma_k\}$ for the pseudoscalar, axial, vector and, axial-vector currents, respectively. After Wick contracting the fields, the correlator can be written as
\begin{equation}
 C_{\Gamma\Gamma'}(t)
 = -\sum_\mathbf{x}\tr\{\gamma^5\Gamma S(\mathbf{x},t)\Gamma'\gamma^5S^\dagger(\mathbf{x},t)\},
\end{equation}
where $S(\mathbf{x},t)=D^{-1}(\mathbf{x},t)$ is the propagator from $(0,0)$ to $(\mathbf{x},t)$.

Numerically we use point-to-all propagators to estimate the two-point functions with the noise-reduction technique described in \cite{Boyle:2008rh} by taking a stochastic average over the volume of the point source. This method has the additional advantage that spin-dilution can be used to calculated all channels with only four inversions. Masses and decay constants for the isotriplet mesons are extracted from the asymptotic behavior of the correlators $C_{\Gamma\Gamma'}$ at large Euclidean time as described in \cite{DelDebbio:2007pz}.

For the two-point correlators we furthermore apply a trick to cancel the backward propagating states \cite{Blum:2001xb}. In practice this allows us to double the extent of the correlators, at the cost of doing an additional set of inversions for each source. With this method we define the propagator as
\begin{equation}
 S_\pm(x,y) = S_A(x,y) \pm S_P(x,y),
\end{equation}
such that the correlator reads
\begin{equation}
 C^\pm_{\Gamma\Gamma'}(t)
 = -\sum_\mathbf{x}\tr\{\gamma^5\Gamma S_\pm(\mathbf{x},t)\Gamma'\gamma^5S^\dagger_\pm(\mathbf{x},t)\}.
\end{equation}
The propagator $S_P(x,y)$ is calculated with periodic boundary conditions in time and $S_A(x,y)$ with anti-periodic boundary conditions in time. With these definitions the first correlator $C^+_{\Gamma\Gamma'}(t)$ gives the forward propagating part from $0$ to $T$ and the second correlator $C^-_{\Gamma\Gamma'}(t)$ gives the backwards propagating part from $2T$ to $T$.

We apply this method because our effective masses in some cases decay too slowly to reach a proper plateau. With the extended correlators the results are significantly better, but a proper plateau might still not be fully reached. For this reason we always calculate the effective masses in two ways.
\begin{enumerate}
 \item Fit a ``possible'' plateau for the effective mass to a constant.
 \item Fit the effective mass to the function $m(t)=m_\infty+ae^{-bt}$ and use $m_\infty$ as the result.
\end{enumerate}
With the extended correlators both methods generally agree within errors. However, the second method usually returns a slightly smaller value, affected by larger uncertainties, such that the difference between the two methods is statistically insignificant.

In the case of the decay constants $f_{PS}$ and $f_{V}$, we use the second method above, as this results in smaller systematic errors from finite volume effects (see Sect.~\ref{sect:fvolume} below). For all other observables we use first method.

All the spectral quantities considered in this work are bare (i.e.~not renormalized) quantities.

\subsubsection*{Baryons}
To extract the mass of the lightest baryon state, we use the following interpolating operator for the spin-1/2 baryon with positive parity.
\begin{equation}
 O_N = \xi^{abc}\{u^aC\gamma_5 d^b\}u^c,
\end{equation}
Here $C=\gamma_0\gamma_2$ is the charge conjugation matrix. The contraction symbol $\xi^{abc}$ is defined in Eq.~\eqref{eq:xi} in Appendix~\ref{sect:sextet}, but the operator is otherwise identical to the one used in QCD.

Because the color contractions are symmetric for this model, the operator formally describes a baryon belonging to the MS flavor representation, as oppose to QCD, where the equivalent operator describes a baryon in the MA flavor representation. This distinction is, however, not important because the operator describes the same state regardless of the flavor representation.

To calculate the baryon correlators we use ordinary point sources, which require 24 inversions for each propagator. The correlator is defined as
\begin{equation}
 C_\pm(t) = \sum_\mathbf{x}\tr\{P_\pm\langle \bar{O}_N(\mathbf{x},t)O_N(0)\rangle\}, 
 \end{equation}
where $P_\pm = \frac{1}{2}(1\pm\gamma^0)$ projects onto positive and negative parity states. Space-time reflection symmetries of the action and the anti-periodic boundary conditions in the temporal direction for the quark fields imply, for zero-momentum correlators, that $C_+(t) = -C_-(T-t)$. Therefore, in order to decrease errors we average correlators in the forward and backward direction
\begin{equation}
 C(t) = \frac{1}{2}[C_+(t) - C_-(T-t)]\, .
\end{equation}
In the following we only consider the positive parity state, as the negative parity state is too noisy to be accurately determined.

\subsubsection*{Gradient flow}
Following \cite{Luscher:2010iy} we consider the gradient flow for the gauge field. Formally the gradient flow is defined through the equation
\begin{equation}
 \partial_t B_\mu(t,x) = D_\nu G_{\nu\mu}(t,x),
\end{equation}
where $G_{\mu\nu}=\partial_\mu B_\nu-\partial_\nu B_\mu + [B_\mu,B_\nu]$ is the field strength tensor and $D_\mu=\partial_\mu+[B_\mu,\cdot]$ is the covariant derivative. The initial condition is $B_\mu(t,x)|_{t=0} = A_\mu(x)$. The fictitious flow time $t$ should not be confused with Euclidean time. For the Wilson action, the discretized flow equation reads
\begin{equation}
\frac{d}{dt} V_\mu(t,x) = -g^2\{\partial_{x,\mu}S_G(V)\}V_\mu(t,x),
\end{equation}
with the initial condition $V_\mu(t,x)|_{t=0}=U_\mu(x)$. While the gradient flow has several advantages, we will simply use it to measure how the lattice spacing changes as a function of the bare parameters. In literature, two related scale-setting observables \cite{Luscher:2010iy,Borsanyi:2012zs} have been introduced.
\begin{align}
 \mathcal{E}(t) &= \langle t^2E(t)\rangle \\
 \mathcal{W}(t) &= t\frac{d\mathcal{E}(t)}{dt}
\end{align}
Here $E=\frac{1}{4}G^a_{\mu\nu}G^a_{\mu\nu}$ is the action of the flowed gauge field. On the lattice we use the symmetric clover discretization for this observable. In both cases the scale setting is performed by choosing a fixed reference value, such that
\begin{align}
 \mathcal{E}(t_0) &= \mathcal{E}_\mathrm{ref}, \\
 \mathcal{W}(w_0^2) &= \mathcal{W}_\mathrm{ref}.
\end{align}
Keeping the reference value fixed, the change in $t_0$ and $w_0$ is then related to the change in lattice spacing, as a function of the change in the bare parameters.

\section{Phase structure}
\label{sect:phases}
In this section we discuss the phase structure of the lattice model. In particular we compare the behavior of the model when simulated in the weak and strong coupling phase, respectively.

\begin{figure}[t]
\centering
 \includegraphics[scale=0.9]{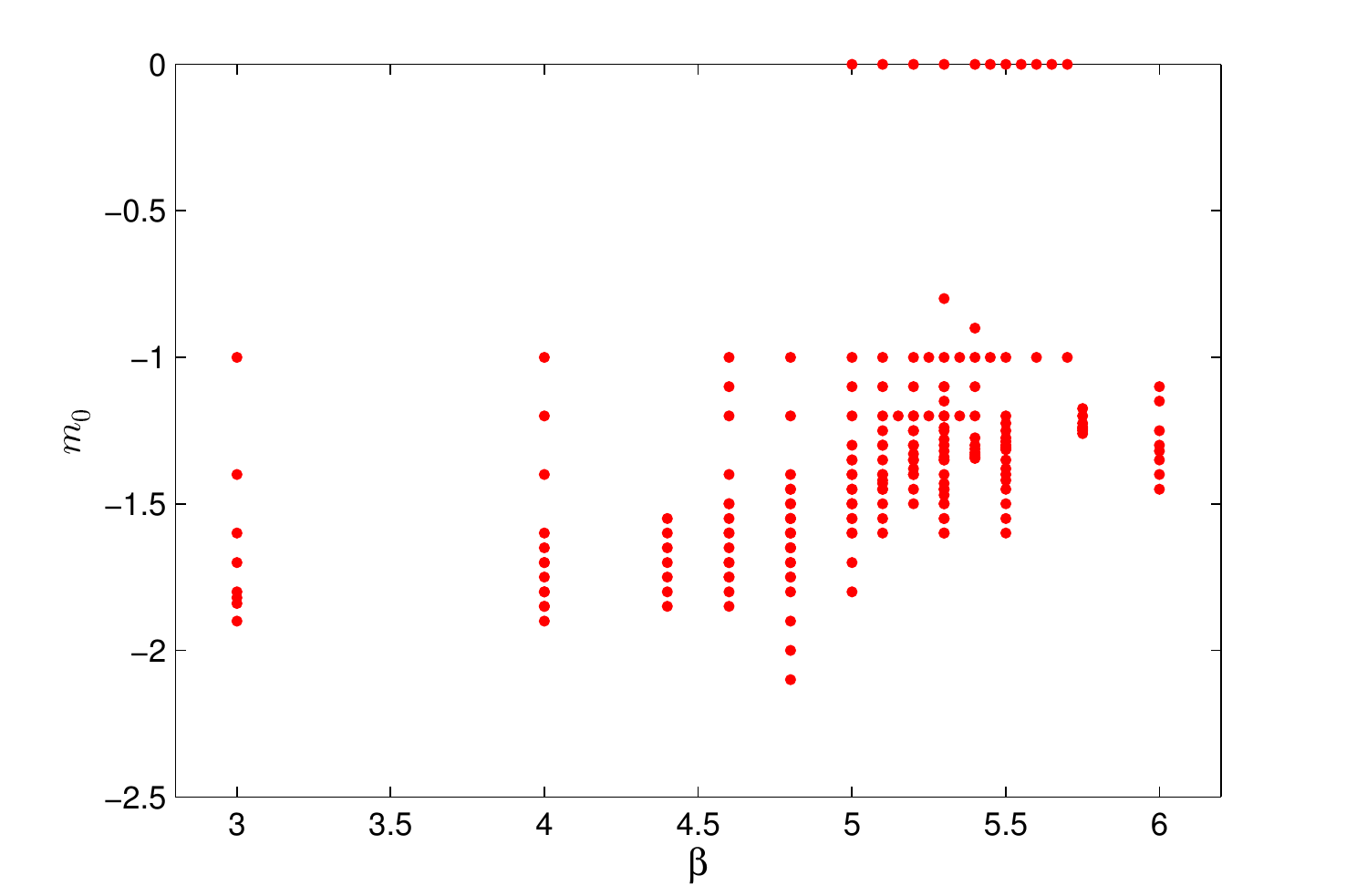}
 \caption{Overview the simulations used for scanning the lattice phase diagram. Some parameters have been simulated twice with different initial configurations (random or unit).}
 \label{fig:simulations}
\end{figure}
\begin{figure}[t]
\centering
 \includegraphics[scale=0.9]{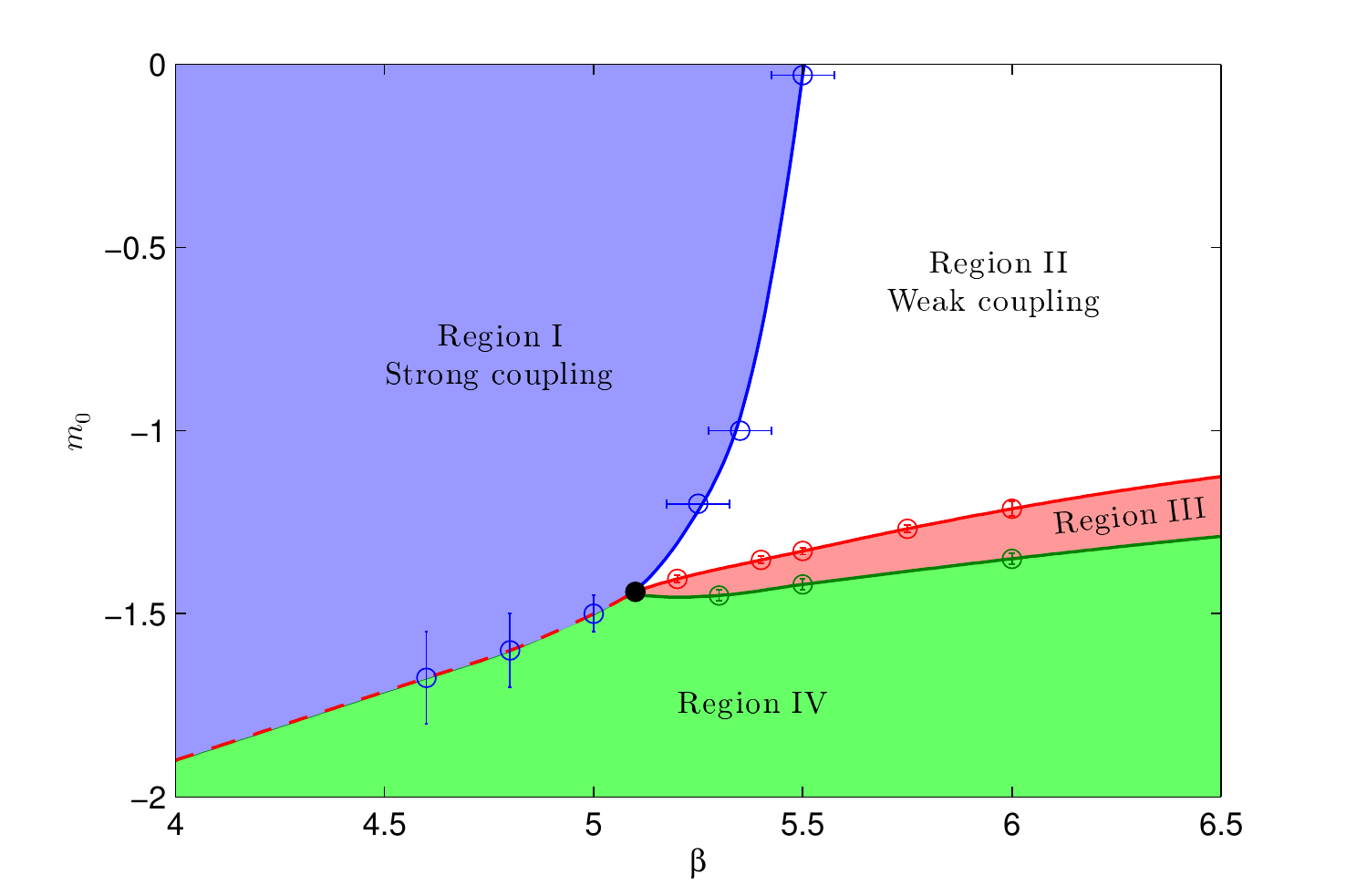}
 \caption{Phase diagram for the lattice model showing four different regions. Region I is a strong coupling bulk phase, region II and III are weak coupling phases with positive and negative PCAC mass, respectively, and region IV is an unphysical artifact region.}
 \label{fig:phase_diagram}
\end{figure}

To properly understand the lattice model and reveal its non-trivial phase structure, we performed an extensive scan, comprising more than 200 simulations, in the parameter space of the bare coupling $\beta$ and the bare quark mass $m_0$. For this scan we use either $8^4$ or $16^3\times32$ lattices, depending on the observable. To check for finite volume effects, some of the simulations have been repeated on $24^3\times48$ lattices. We show in Fig.~\ref{fig:simulations} an overview of the simulations used for scanning the $(\beta,m_0)$ plane.

From the scan we are able to determine different phases of the lattice model. In Fig.~\ref{fig:phase_diagram} we show the resulting phase diagram, where we identify four different regions separated by either first order transitions (dashed lines) or continuous transitions (solid lines).

\begin{figure}[t]
\centering
 \includegraphics[scale=0.85]{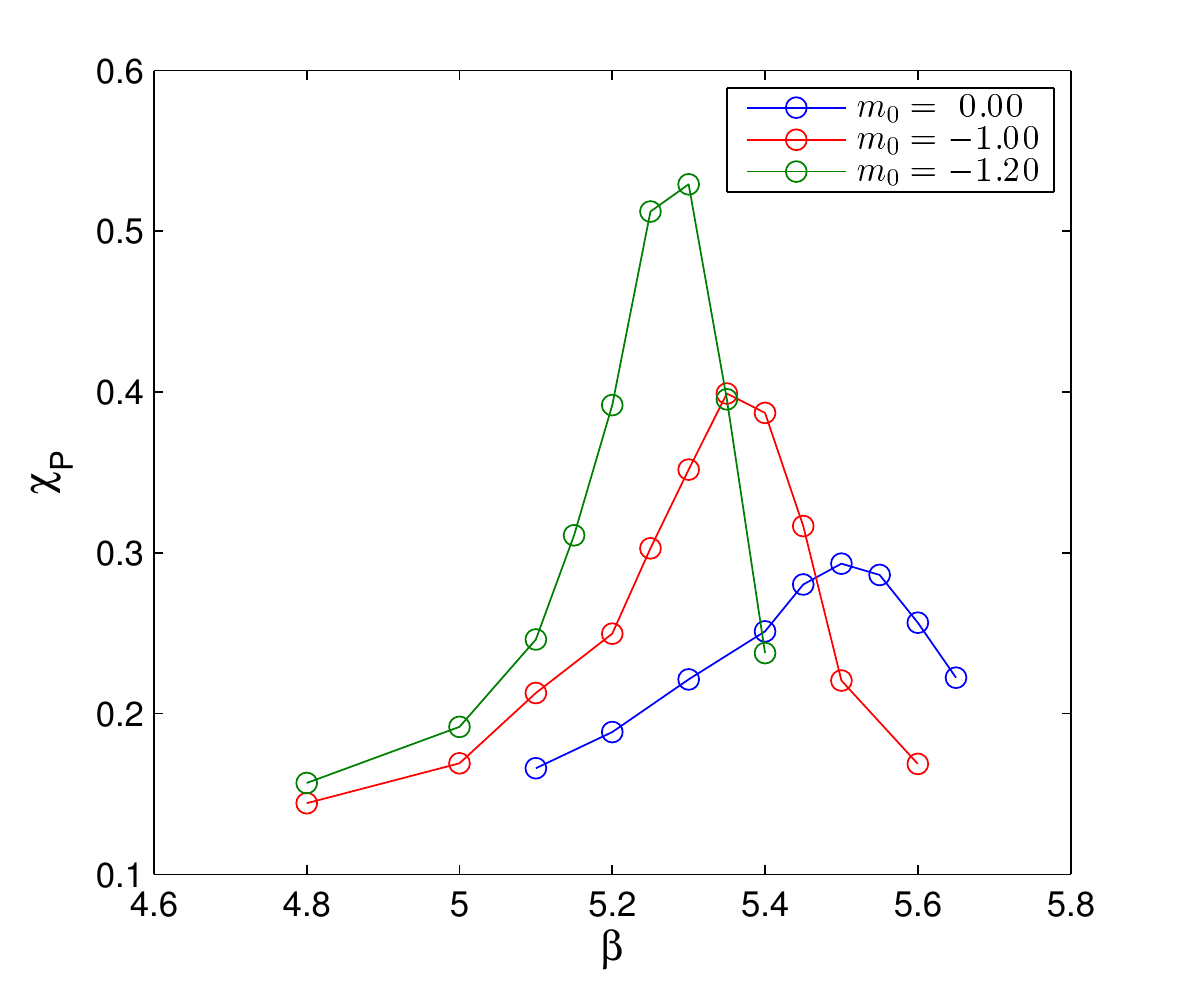}
 \caption{Plaquette susceptibility for positive PCAC mass at three different values of $m_0$. The location of peak determines the crossover transition line between the ``weak coupling phase'' and the ``strong coupling phase'' (solid blue line in Fig.~\ref{fig:phase_diagram})  .}
 \label{fig:susc}
\end{figure}

On the figure, the solid blue line is a continuous crossover identified by looking at the peak of the plaquette susceptibility, as a function of the bare coupling, in the region of positive PCAC mass. This line separates the ``weak coupling phase'' from the ``strong coupling phase''.
The plaquette susceptibility is defined as
\begin{equation}
\chi_P = \frac{\partial\langle P \rangle}{\partial \beta}.
\end{equation}
We plot in Fig.~\ref{fig:susc} the plaquette susceptibility for three different values of the bare mass. The susceptibility shows a clear peak, whose maximum value increases when decreasing the bare masse, i.e. when approaching the limit of vanishing quark mass.
The maximum of the plaquette susceptibility seems to be volume independent, which indicates that the transition is a smooth crossover which becomes sharper at lower quark masses.

\begin{figure}
\centering
 \includegraphics[scale=0.8]{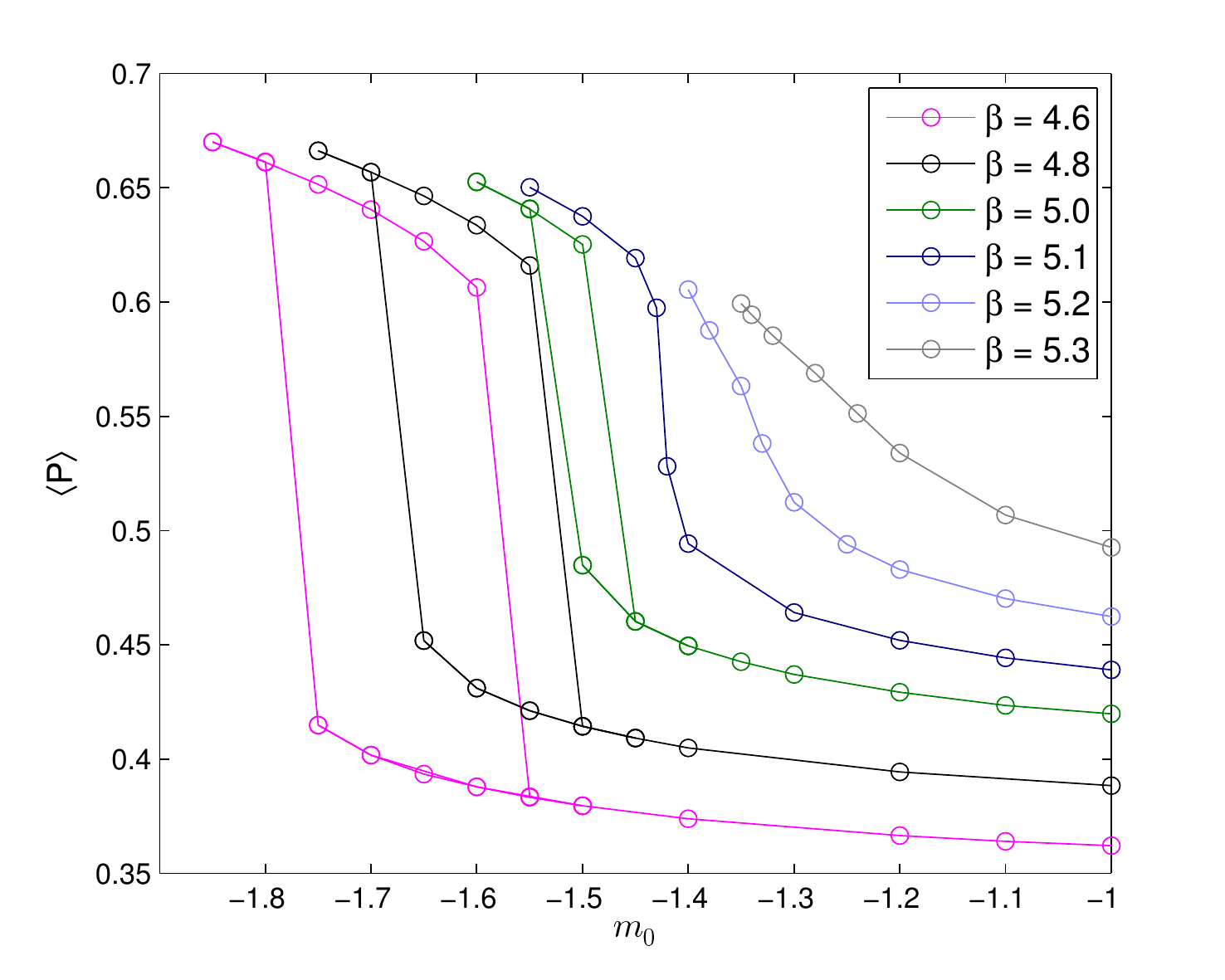}
 \caption{Average value of the plaquette $\langle P\rangle$. At strong coupling a first order transition is observed as a jump in the plaquette average.}
 \label{fig:plaq}
\end{figure}

At strong coupling a line of first order transition is present (the dashed red line in Fig.~\ref{fig:phase_diagram}) with an endpoint around $\beta\approx 5.1$ as indicated by the black point.
This transition line is identified by discontinuous jumps in the average value of the plaquette. We show in Fig.~\ref{fig:plaq} an overview of the average plaquette for different values of $\beta$ in the strong and weak coupling phases. A discontinuity is clearly visible indicating the first order transition.

\begin{figure}[H]
 \centering
 \includegraphics[scale=0.8,trim={0 20mm 0 20mm},clip]{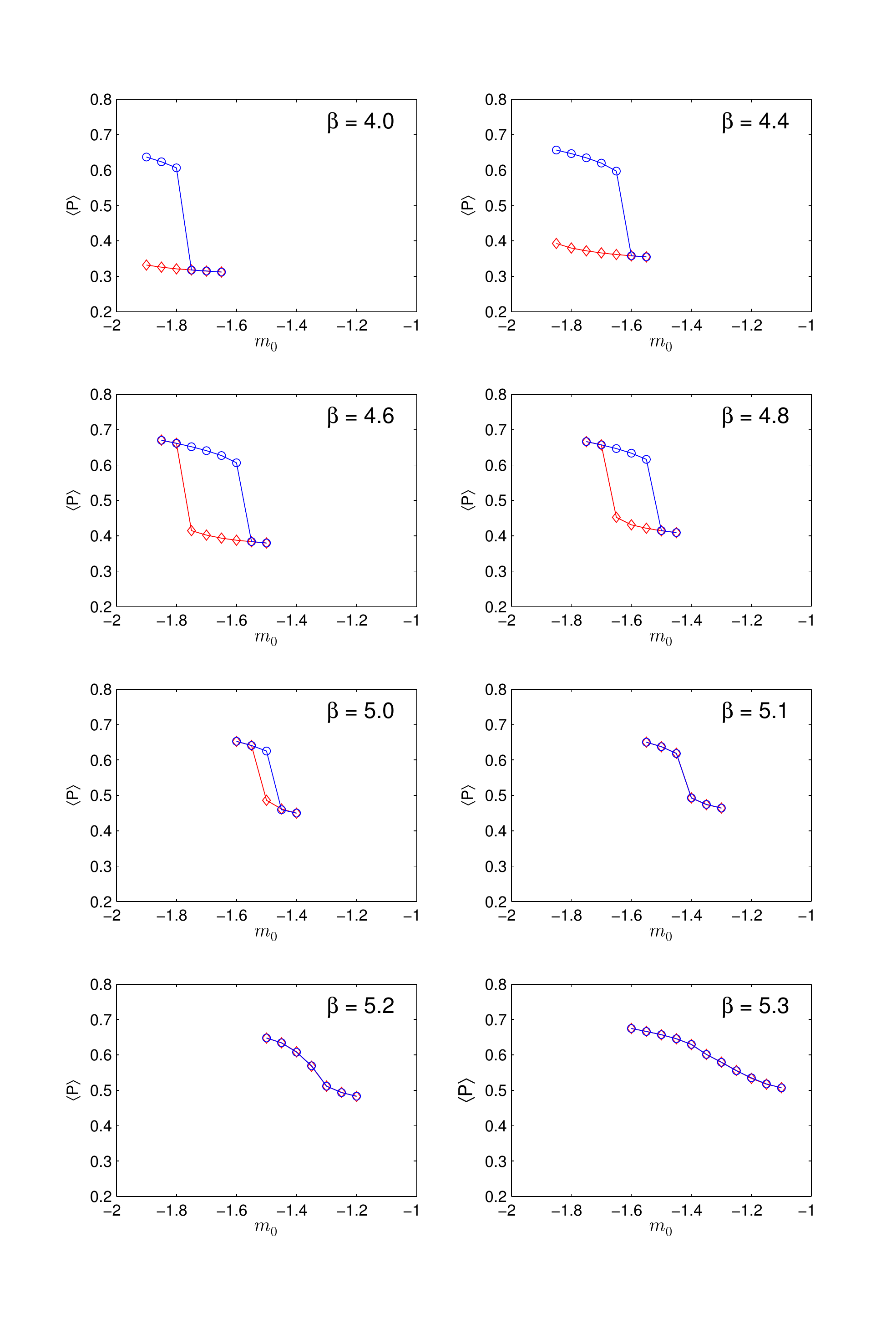}
 \vspace{-8mm}
 \caption{Hysteresis cycles around the first order phase transition. The red (blue) points indicate simulations started from a random (unit) configuration. At strong coupling we are unable to run simulations for bare masses below $m_0\approx-1.9$. As we approach the weak coupling phase, the first order transition disappears.}
 \label{fig:hysteresis}
\end{figure}

At strong coupling, hysteresis cycles appear in simulations started from different initial configurations, as shown in Fig.~\ref{fig:hysteresis}. All data points on this figure have been obtained from independent simulations started from either a unit configuration (blue points) or a random configuration (red points). At very strong coupling we are unable to run simulations for bare masses below $m_0\approx-1.9$, which is why the hysteresis cycles in top most panel are not closed.
As we approach the weak coupling region, the hysteresis cycles disappear, as seen in the panel at the bottom. At $\beta=5.2$ one can still observe a small signal for the transition, while at $\beta=5.3$ it is absent.

For the study of hysteresis cycles, we used small $8^4$ lattices, since the presence of strong metastabilities on large lattices makes it difficult to perform numerical simulations across the transition. To check for the persistence of the first order transition, we have repeated the simulation at $\beta=5.0$ and $m_0=-1.5$ on the three volumes $8^4$, $16^3\times32$ and $24^3\times48$, and confirmed that the average plaquette value is independent of the lattice volume.
As an illustration, in Fig.~\ref{fig:plaq} several of the points on the hysteresis cycles are from a small $8^4$ volume and one of the larger volumes. The difference in the average plaquette are not visible at this scale.

The line of first order transition continues as a continuous transition line (solid red line in Fig.~\ref{fig:phase_diagram}) in the weak coupling regime. This line is identified as the line where the PCAC mass vanishes, when approaching from positive bare mass.

Finally we identify one more line (solid green line in Fig.~\ref{fig:phase_diagram}) in the phase diagram. This line is defined as the point where the slope of the PCAC mass changes sign. We discuss the significance of this line in Subsection~\ref{sect:pd-spec}.

The outlined phase diagram shares a number of features with the phase diagram of the SU(2) gauge theory with two Dirac fermions in the adjoint representation, first studied in \cite{Hietanen:2008mr}, which is a model known to be inside the conformal window~\cite{Catterall:2007yx,Hietanen:2008mr,DelDebbio:2008zf,DelDebbio:2009fd,Hietanen:2009az,Catterall:2009sb,Bursa:2009we,DelDebbio:2010hx,DelDebbio:2010hu,Bursa:2011ru,DeGrand:2011qd,DelDebbio:2015byq,Rantaharju:2015cne,Rantaharju:2015yva}.

The common features includes a first-order phase transition at strong coupling, with an endpoint which then becomes a continuous transition line where the quark mass vanishes. We will show below that similar behaviors of the spectral quantities both at strong and weak coupling are also observed.

In simulations with many fundamental flavors, the same first-order phase transition at strong coupling is also observed \cite{Nagai:2009ip}, indicating that this might be a general feature for Wilson fermions for models inside or close to the conformal window.

\subsection{Spectrum}\label{sect:pd-spec}
Having understood the phase structure of the discretized lattice model, we now turn our attention to several interesting observables, starting from the light meson spectrum.

\begin{figure}
\centering
 \includegraphics[scale=0.55,trim={5mm 0 10mm 0},clip]{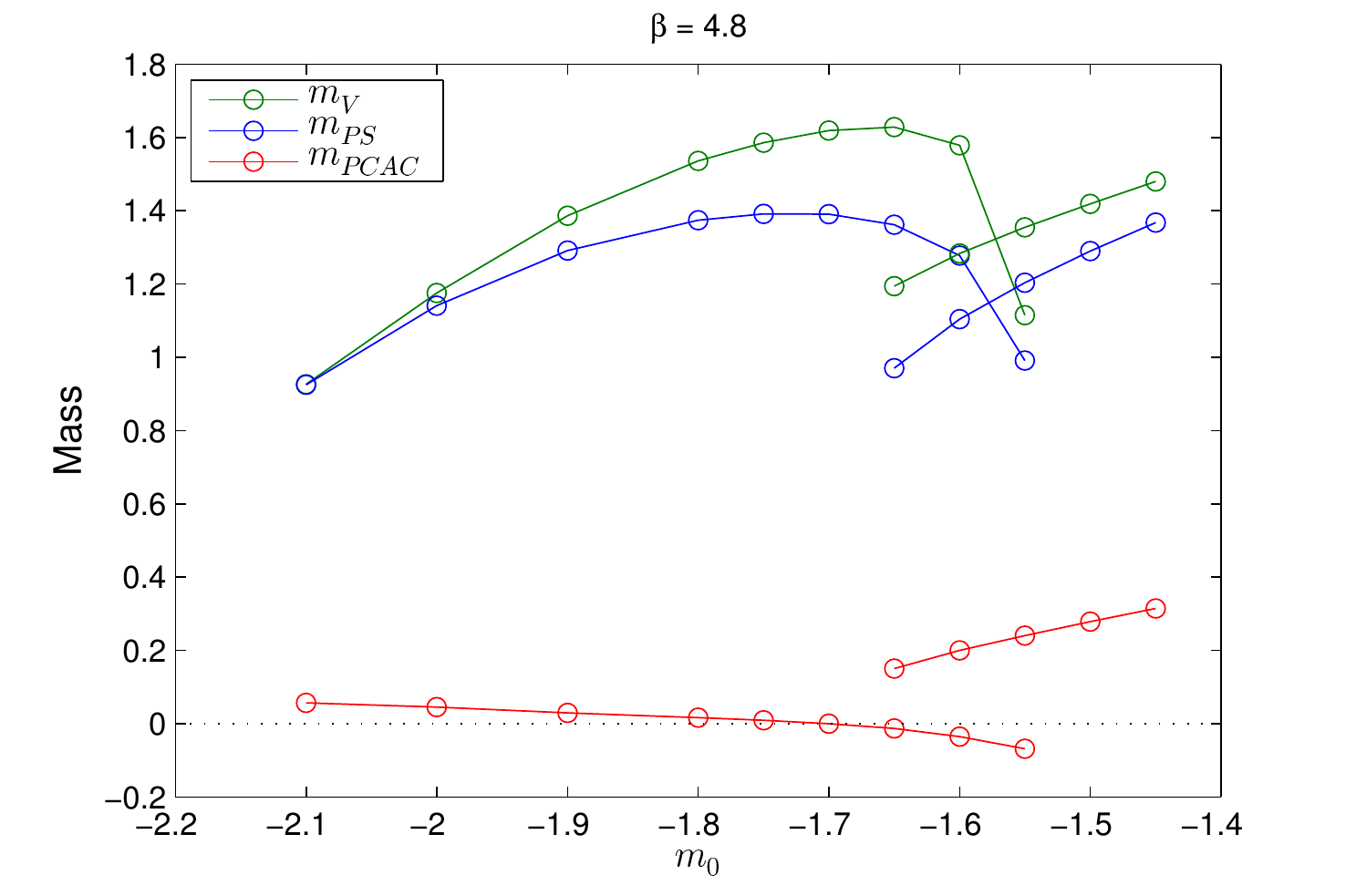}
 \includegraphics[scale=0.55,trim={5mm 0 10mm 0},clip]{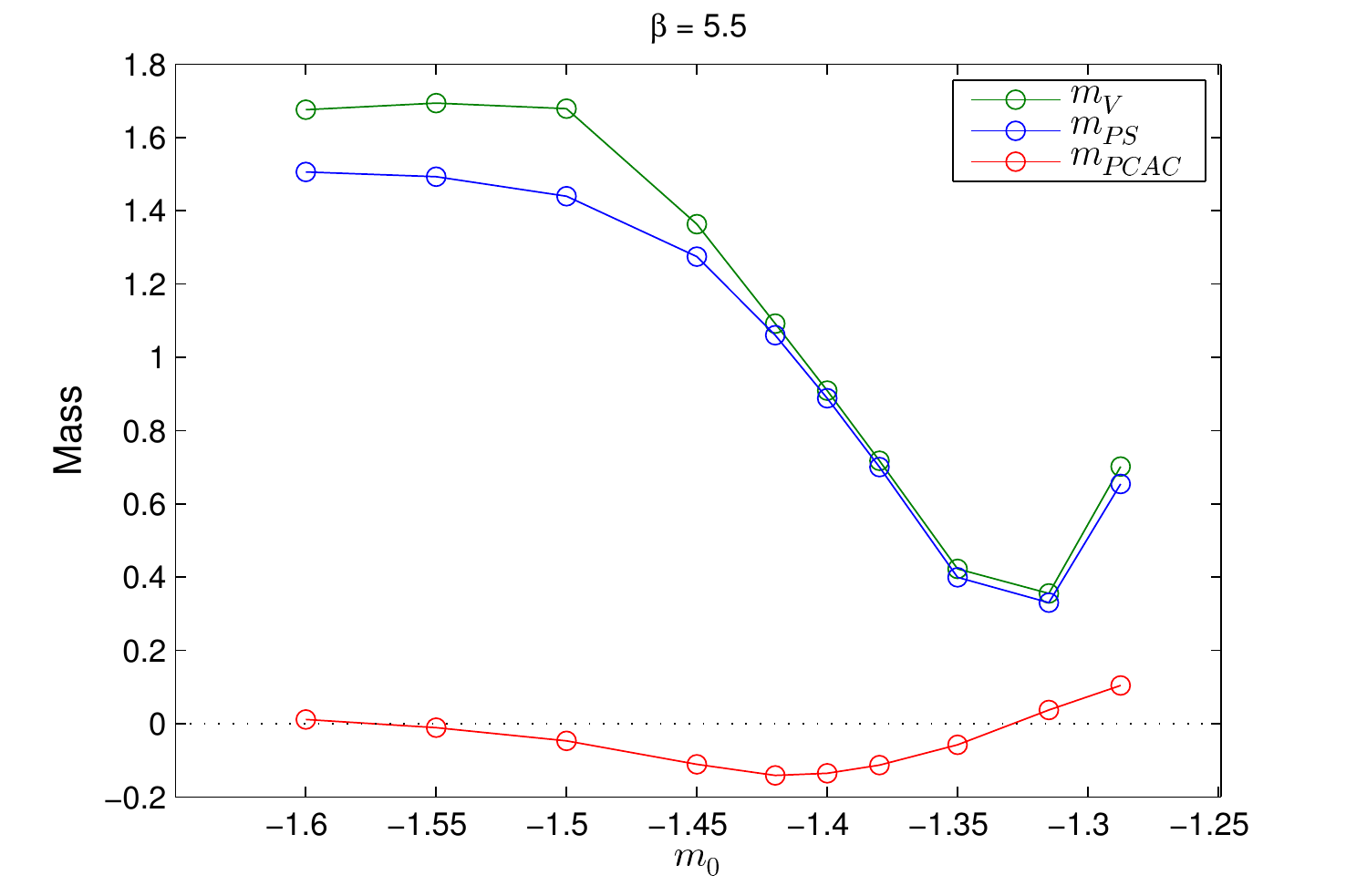}
 \caption{\textit{Left}: Behavior of $m_V$, $m_{PS}$, and $m_{PCAC}$ when crossing the first-order transition at $\beta=4.8$. The lines extending furthest the right (left) are simulations in Region I (Region IV). We observe that the slope of the PCAC mass has opposite sign in Region IV. \textit{Right}: At $\beta=5.5$ there is a continuous transition between Region II, III and IV, and we clearly see that the slope of the PCAC mass changes sign around $m_0\approx-1.42$.}
 \label{fig:1storder}
\end{figure}

\begin{figure}
\centering
 \includegraphics[scale=0.55,trim={5mm 0 10mm 0},clip]{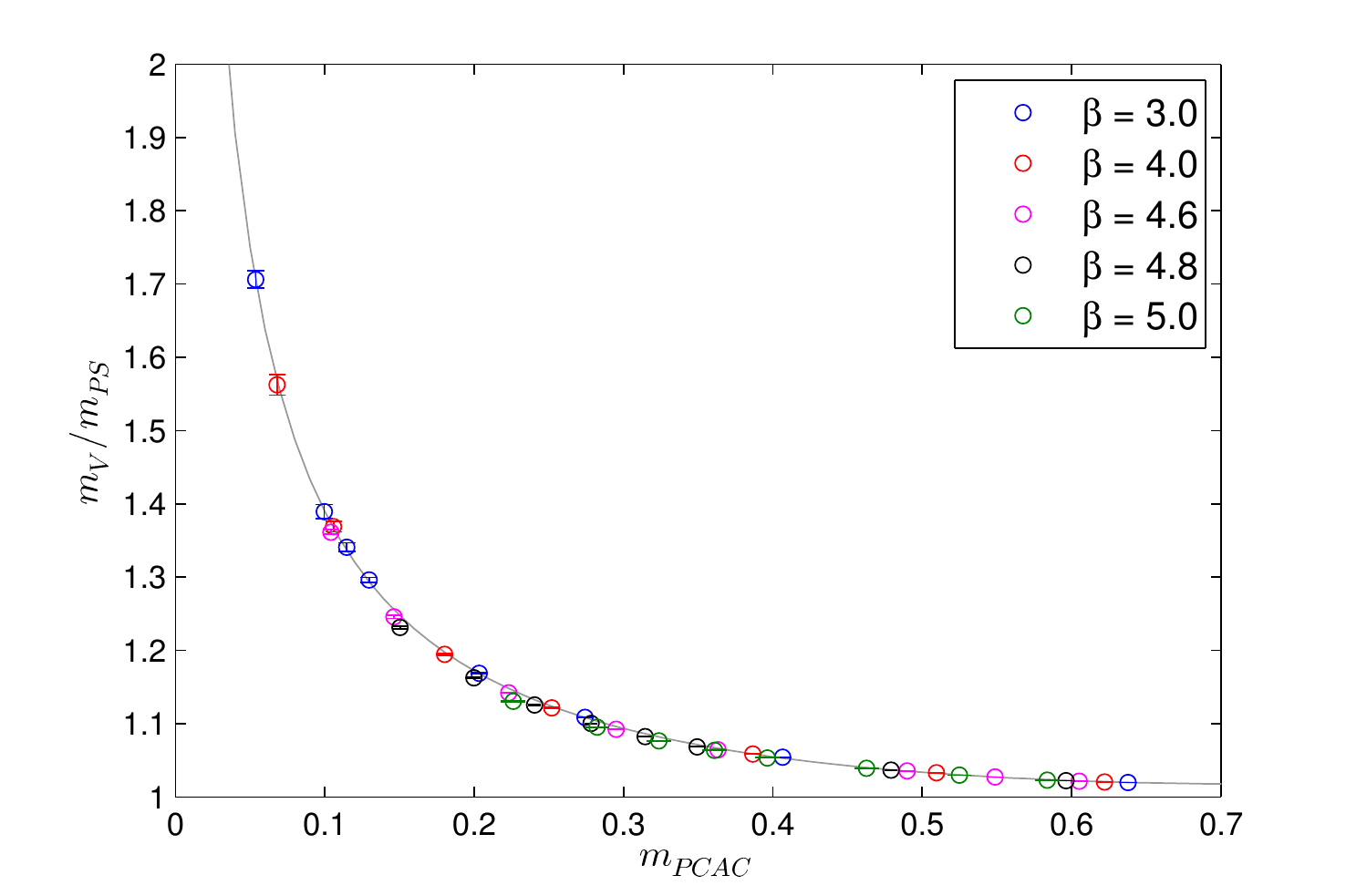}
 \includegraphics[scale=0.55,trim={5mm 0 10mm 0},clip]{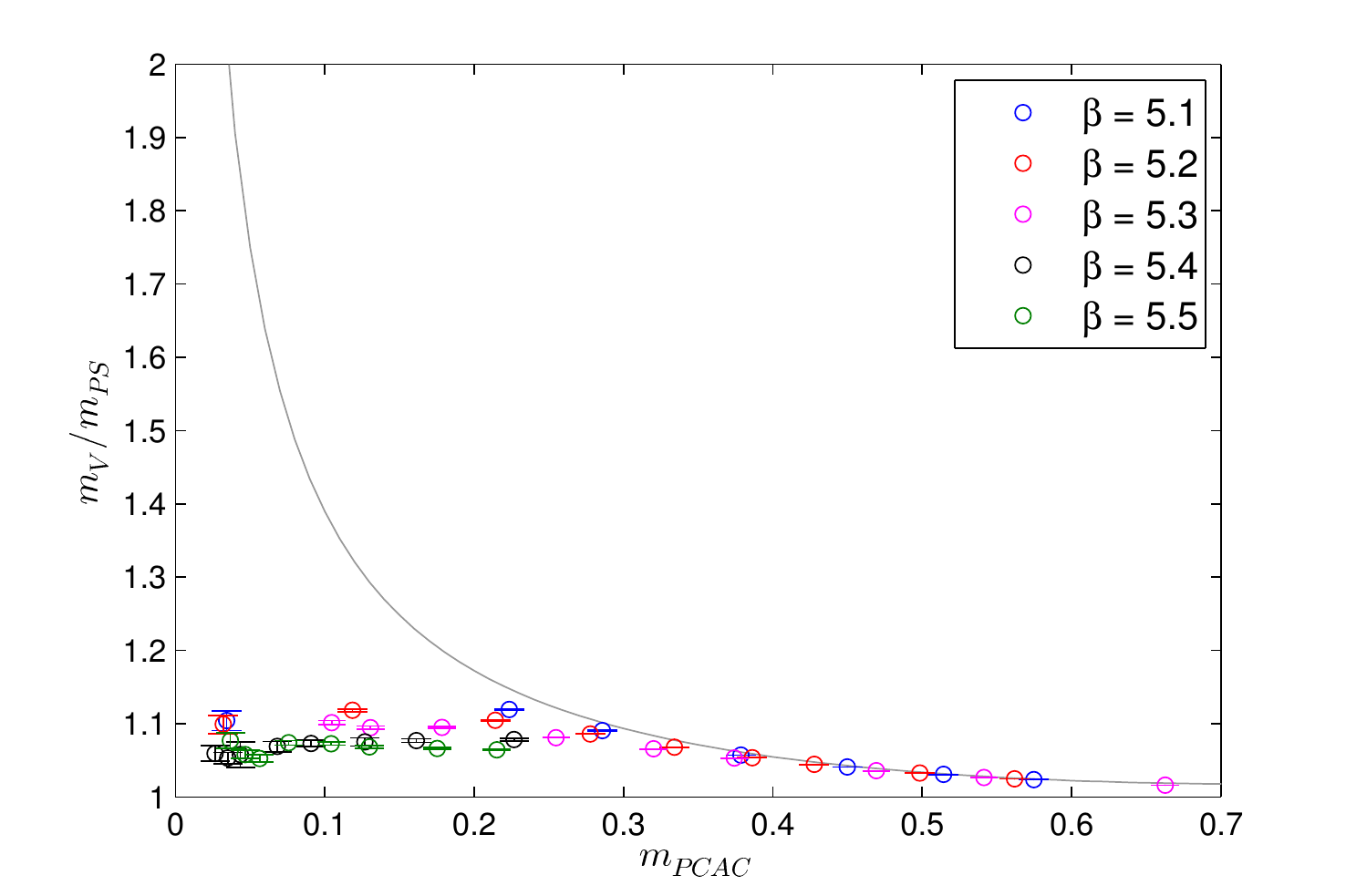}
 \caption{The ratio $m_V/m_{PS}$ as a function of the PCAC mass and inverse coupling $\beta$. \textit{Left}: Inside the bulk phase (Region I) the ratio diverges in the chiral limit, as expected from chiral symmetry breaking. The gray line is a fit to the data at $\beta=3.0$ and $\beta=4.0$. \textit{Right}: In the weak coupling phase (Region II) the ratio is constant in the chiral limit, as expected in a conformal model.}
 \label{fig:mvkmps}
\end{figure}

In Fig.~\ref{fig:1storder} we show how the PCAC, pseudoscalar, and vector mass change as a function of the bare quark mass across different phases. In the strong coupling region at $\beta=4.8$ (left panel), where a first-order transition is visible, we observe a hysteresis region and discontinuous jumps for the masses. In the figure, the lines extending furthest to the left (right) are from simulations started from a unit (random) configuration. In particular the PCAC mass jumps from a positive to a negative value across the transition, implying that the chiral limit cannot be reached in the strong coupling Region I.

In the right panel of Fig.~\ref{fig:1storder} we show the same observables for $\beta=5.5$ in the weak coupling phase. As the transitions separating the different regions in phase space are all continuous at this weaker coupling, we observe no jumps in the measured observables. The PCAC mass smoothly approaches zero, becomes negative before its slope changes sign around $m_0\approx-1.42$ to increase again until it vanishes for a second, more negative, value of the bare quark mass.
The border between Region III and IV is defined at the point when the slope of the PCAC mass changes sign.
This defines Region III as the part of the phase diagram at weak coupling where the chiral transition line can be approached from negative PCAC masses and where we observe a qualitatively similar behavoir for the meson masses when compared to the positive PCAC mass case in Region II.

In the rest of this paper we will focus our attention only on Region I and II, where one can approach the chiral line from positive PCAC mass values.

In Fig.~\ref{fig:mvkmps} we show how the ratio $m_V/m_{PS}$ changes when moving from Region I (the strong coupling phase) to Region II (the weak coupling phase) as a function of the PCAC mass.
In the strong coupling phase (left panel of the figure), where hysteresis exists, we show the ratio of masses as obtained from simulations starting from random configurations.
This ratio clearly increases towards the chiral limit and we observe a clear splitting between the vector and pseudoscalar meson masses, which is consistent with the expectation of chiral symmetry breaking. This is not surprising, as lattice artifacts always break chiral symmetry at strong coupling.
We also note that, since the first order transition becomes stronger and the metastable states are more stable at stronger couplings, we are able to reach much lower positive PCAC masses for stronger couplings.

When moving to weak couplings in Region II (right panel) we observe that the ratio becomes almost constant towards the chiral limit and the two states remain almost degenerate over the entire range of quark masses investigated here.
Most of the results in the right panel of Fig.~\ref{fig:mvkmps} are obtained on a $16^3\times 32$ lattice volume, but we also include our data from large volume simulations at $\beta=5.4$ and $\beta=5.5$ which we use later in the paper for a more detailed analysis of the spectrum at weak coupling. There is a small difference in the ratios obtained over the two volumes, due to finite size effects.

If persisting to arbitrarily small quark masses and weak couplings, this flat behavior would indicate the absence of spontaneous chiral symmetry breaking and it would be consistent with the expected hyperscaling behavior of an infrared conformal model \cite{DelDebbio:2010hu,DelDebbio:2010jy,DelDebbio:2010ze}. This drastic change in the qualitative behavior of the model shows a sharp transition between the strong and the weak coupling phases.

\subsection{Gradient flow}
As our scan of the parameter space includes many different values of the bare coupling, in order to understand how the lattice spacing changes, we measure the scale setting observables introduce in Section~\ref{sec:observables}. The reference values used in this work are $\mathcal{E}_\mathrm{ref}=\mathcal{W}_\mathrm{ref}=1$ and $0.15$, which, although different from the usual values used in QCD, give a reasonable definition in the sextet model. We choose two, quite different values, to show that any reasonable choice of the reference values gives the same qualitative behavior. 

\begin{figure}
 \centering
 \includegraphics[scale=0.54,trim={6mm 0 6mm 0},clip]{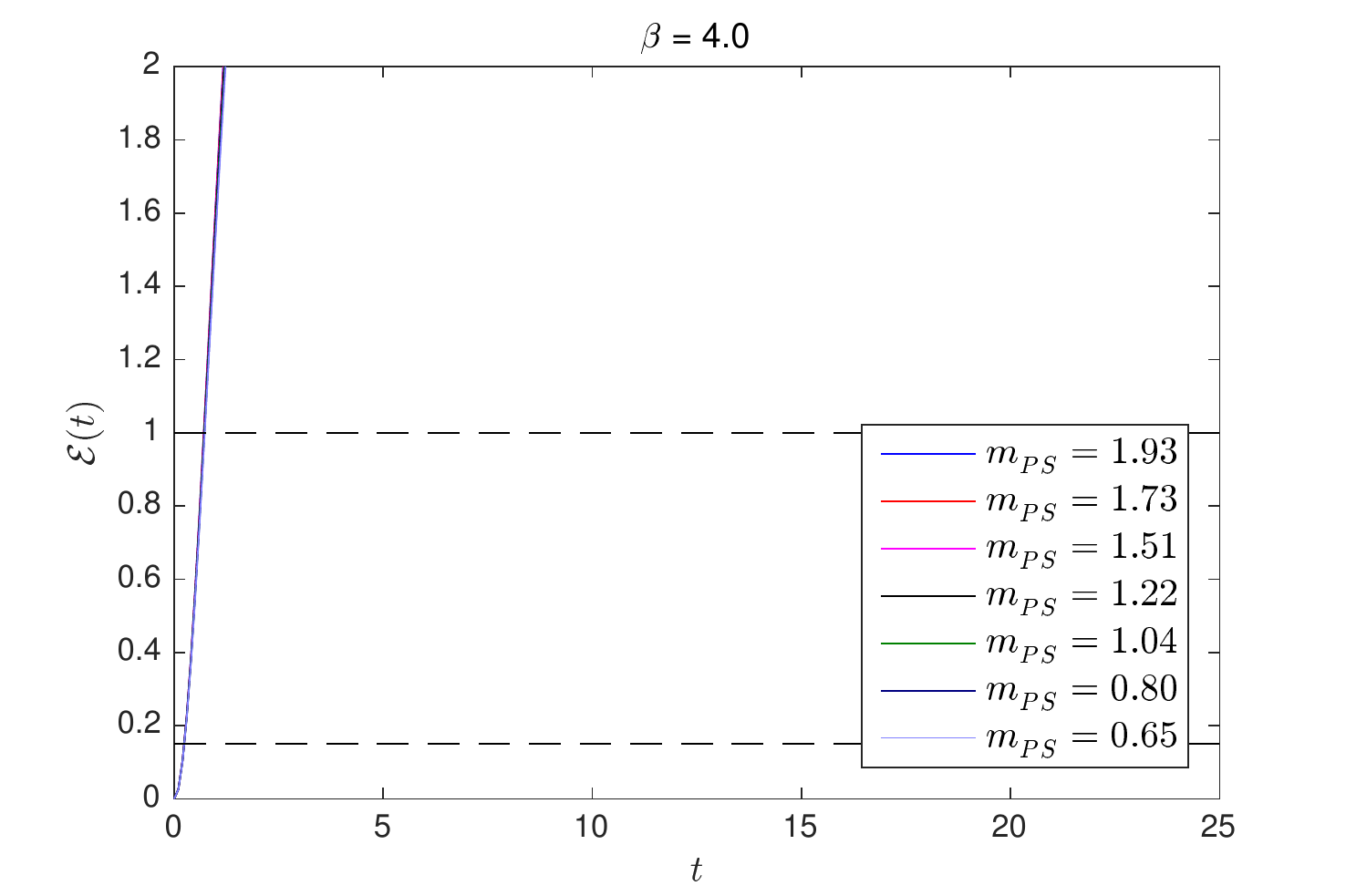}
 \includegraphics[scale=0.54,trim={6mm 0 6mm 0},clip]{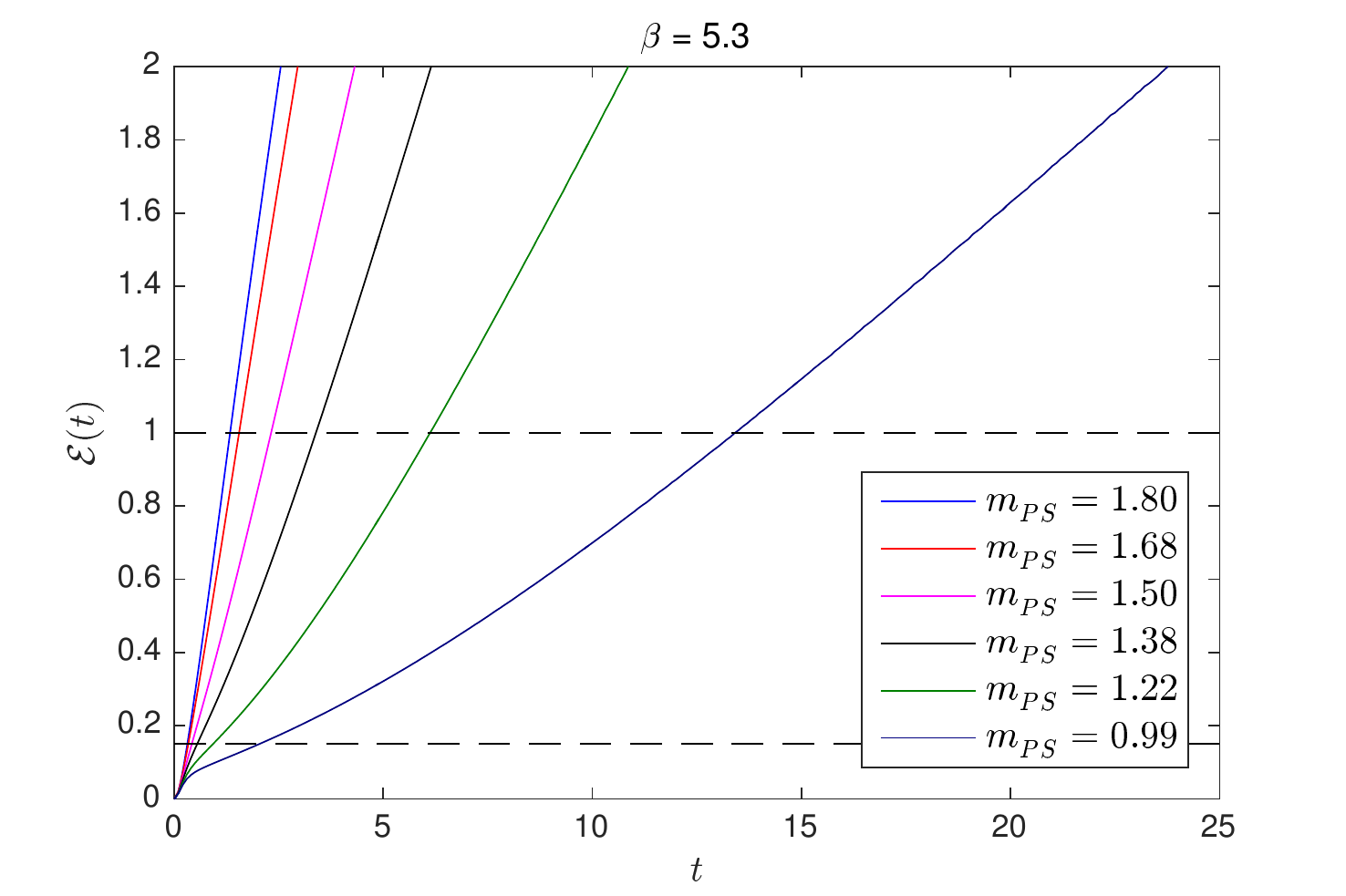}
 \caption{Determination of the $t_0$ observable for two different bare couplings. The left panel at $\beta=4.0$ is in the strong coupling region (Region I), while the right panel at $\beta=5.3$ is in the weak coupling region (Region II). The observable has a negligible quark-mass dependence in Region I, but a very strong quark-mass dependence in Region II. The dashed horizontal lines are the reference values $\mathcal{E}_\mathrm{ref}=1$ and $\mathcal{E}_\mathrm{ref}=0.15$. The quark mass dependence in the two regions is not influenced by the choice of the reference value.}
 \label{fig:flow}
\end{figure}

In Fig.~\ref{fig:flow} we show $\mathcal{E}(t)$ and the determination of $t_0$ for a range of bare masses at $\beta=4.0$ in the strong coupling phase (left panel) and $\beta=5.3$ in the weak coupling phase (right panel).
As before, results in the strong coupling phase are obtained from simulations starting from a random gauge configuration.
From the comparison between the two, it is evident that, while $\mathcal{E}(t)$ in the strong coupling region shows an almost negligible small quark-mass dependence down to the smallest quark mass which is possible to reach, at weak coupling a very strong quark-mass dependence is observed.
In particular the value of $t_0$ seems to diverge when approaching the chiral limit.
A very similar behavior is also observed for the $w_0$ observable.

\begin{figure}
 \centering
 \includegraphics[scale=0.54,trim={6mm 0 6mm 0},clip]{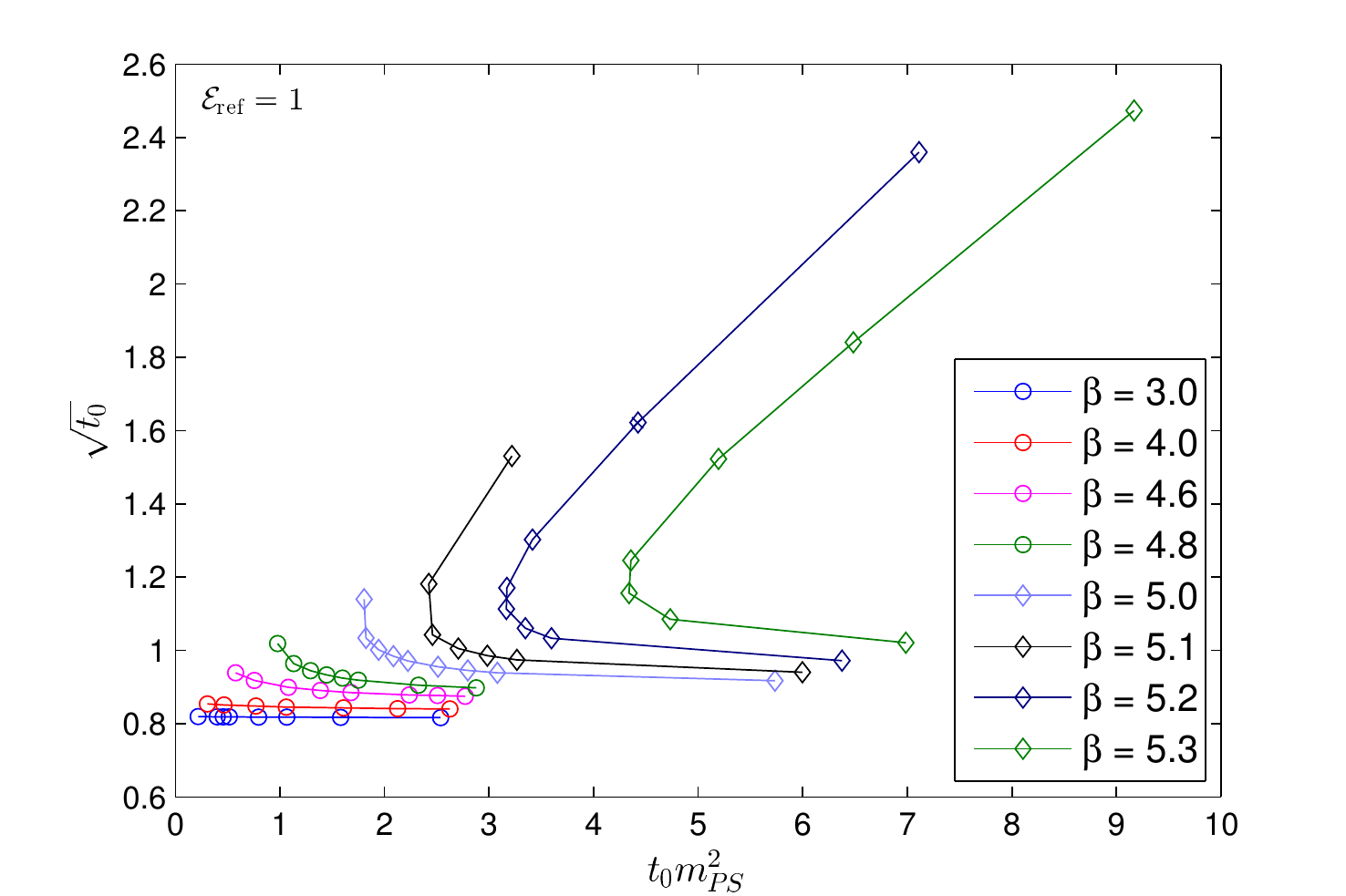}
 \includegraphics[scale=0.54,trim={6mm 0 6mm 0},clip]{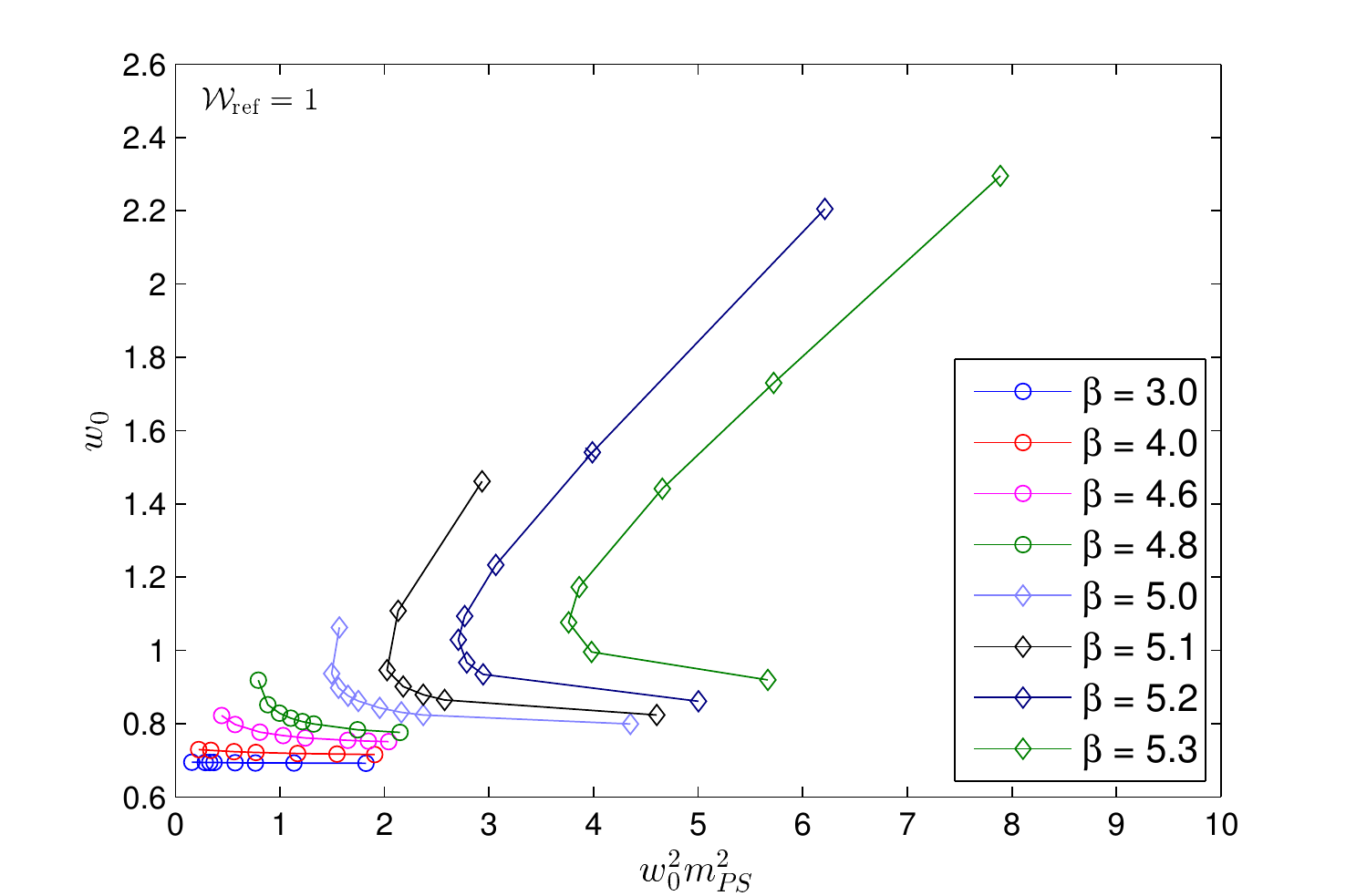}
 \includegraphics[scale=0.54,trim={6mm 0 6mm 0},clip]{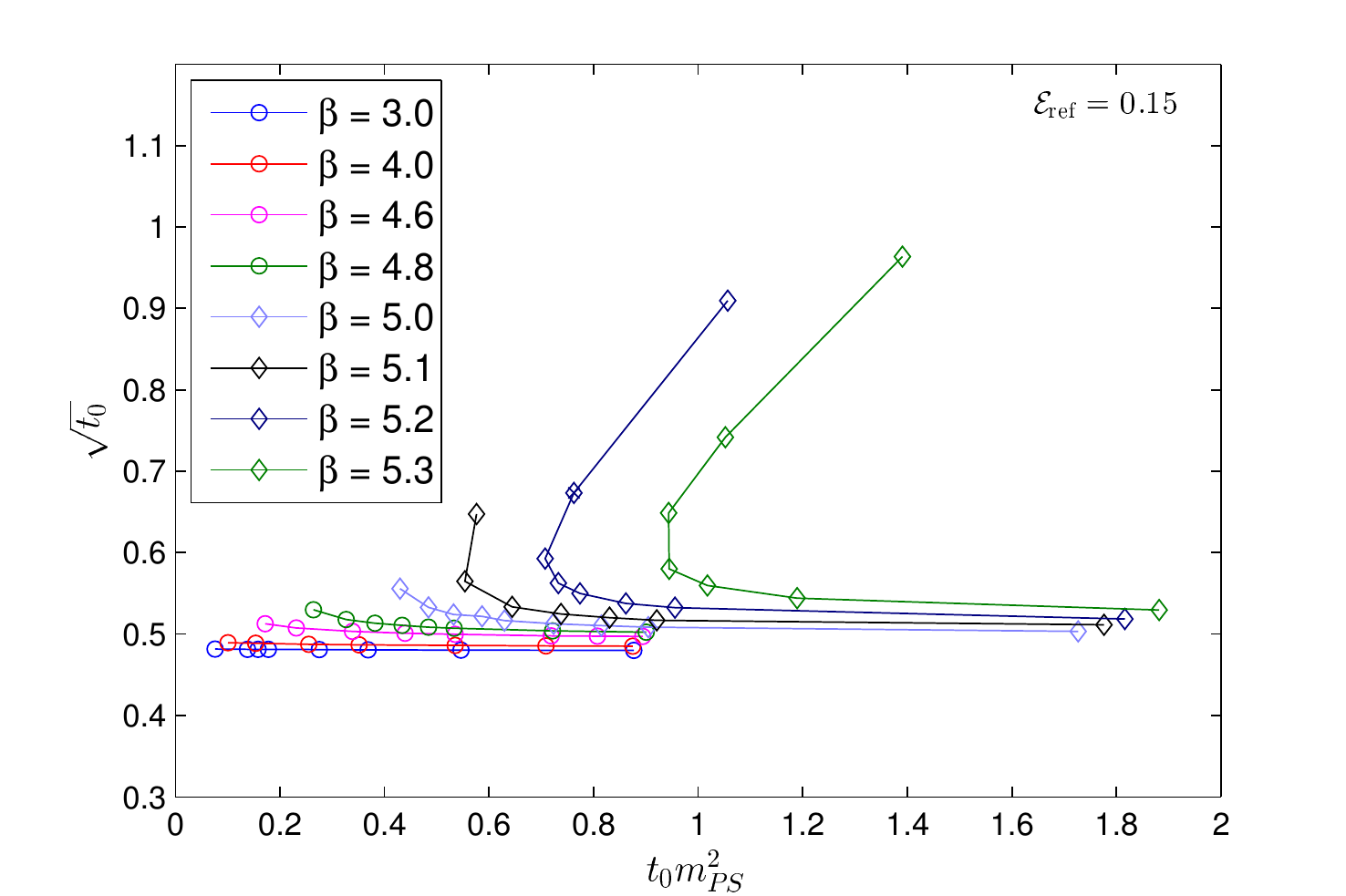}
 \includegraphics[scale=0.54,trim={6mm 0 6mm 0},clip]{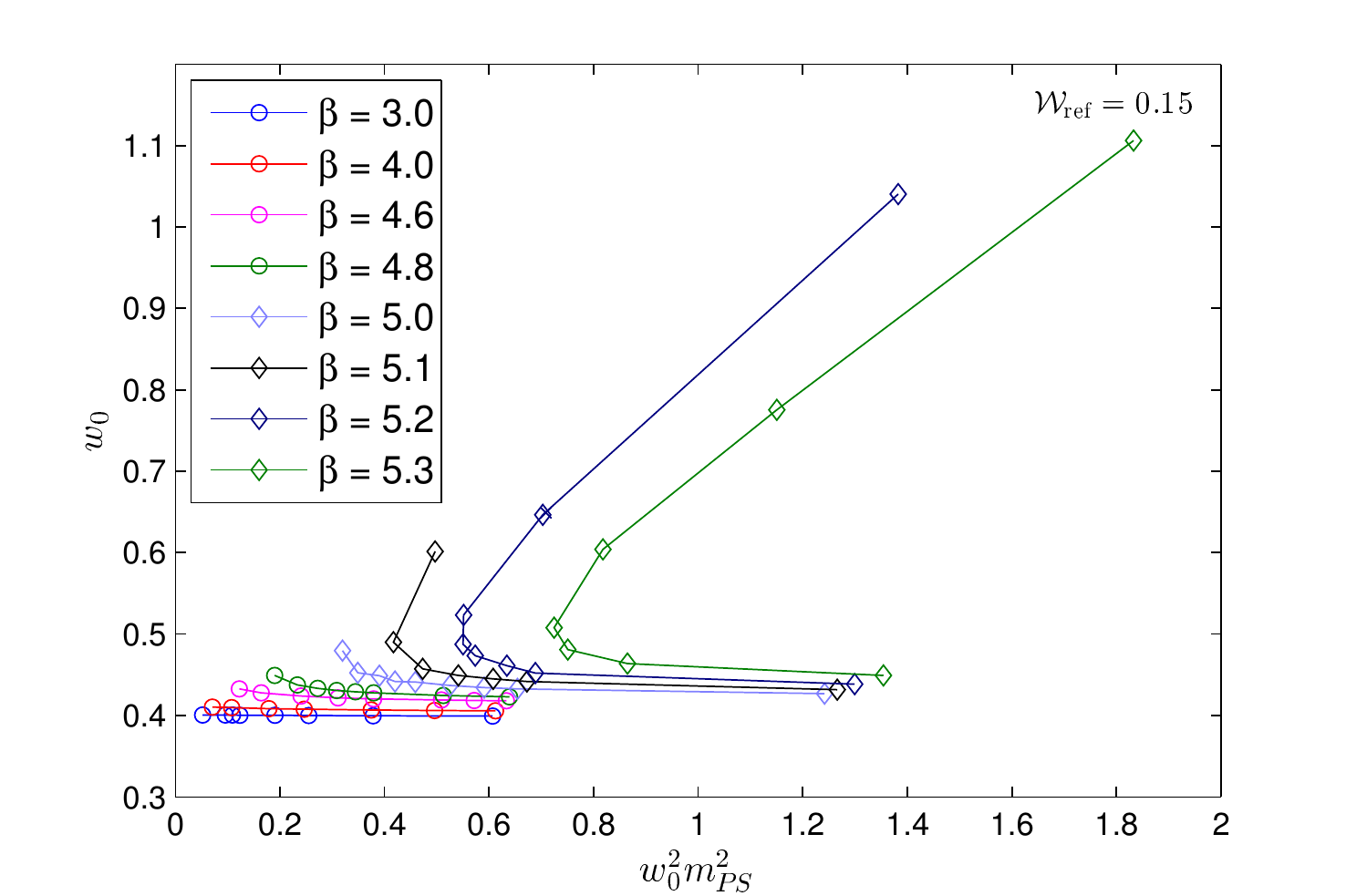}
 \caption{Behavior of $t_0$ and $w_0$ as a function of  $t_0m_{PS}^2$ (left panel) and $w_0^2m_{PS}^2$ (right panel) respectively. The observed turnaround is caused by the very strong quark mass dependence of both $t_0$ and $w_0$ in the weak coupling region. Finite size effects are negligible on both $t_0$ and $w_0$, as can be seen in the figure by looking a the cross symbols which are obtained on a larger lattice volume of $24^3\times 48$. The top and bottom rows are the results for two different choices of the reference scale.}
 \label{fig:w0t0mps}
\end{figure}

This divergent behavior is shown in Fig.~\ref{fig:w0t0mps} where we plot $\sqrt{t_0}$ and $w_0$ as a function of $t_0m_{PS}^2$ (left panel) and $w_0^2m_{PS}^2$ (right panel). In the strong coupling region the extrapolation towards the chiral limit is mild, as expected from previous studies in QCD. In contrast, in the weak coupling phase the strong quark-mass dependence of the two quantities $w_0$ and $t_0$ is clearly seen as a turnaround of the curves in Fig.~\ref{fig:w0t0mps} for $\beta\gtrsim5.0$.
This indicates that $t_0$ and $w_0^2$ are diverging faster than $1/m_\pi^2$ when approaching the chiral limit.

The observed behavior is clearly in contrast to the expectations in a chirally broken model \cite{Bar:2013ora}. We also observe that the measure of $t_0$ and $w_0$ is not affected by finite-volume effects.
This is illustrated in Fig.~\ref{fig:w0t0mps}, where for one of the masses at $\beta=5.2$ we included a comparison with a larger $24^3\times 48$ volume (the cross symbols in the figure).

\section{Weak coupling phase}
\label{sect:wcp}
Having studied the lattice phase diagram of the model, we focus here on the weak coupling phase and, in particular, on the light quark-mass region. The qualitative features which emerged from the phase diagram indicate that in the weak coupling phase no clear signs of chiral symmetry breaking are visible.
The aim of this section, is to understand in more detail if the model shows signs of conformal scaling instead of chiral symmetry breaking in the massless limit. 
We know from the numerous previous numerical studies of near conformal models that the many systematic effects, which are necessarily present in a lattice simulations, can easily obscure the expected behavior, both in the case of IR conformal and chirally broken models. 
One should therefore always try to control all the systematics in the best possible manner.

The strategy that we use, is to look at the infinite volume limit of all observables considered, i.e. at each given quark mass we use a large enough volume so that finite-volume corrections are under control.
A preliminary comment is in order. In the weak coupling phase and at light masses, the sextet model considered here is known to be affected by a severe topological freezing problem. We report our results for the topology in Appendix~\ref{sect:top}, see also e.g. \cite{Fodor:2014pqa} for simulations with staggered fermions.
The weak coupling results presented in this section for light masses are obtained from simulations at zero topological charge. We consider the fixed topology as an additional source of finite volume corrections.

All the data used in this section, were obtained at bare coupling $\beta=5.4$ on lattice volumes of either $24^3\times 40$ or $32^3\times 48$ for the two lightest quark masses.

\begin{figure}
\begin{center}
 \includegraphics[scale=0.50]{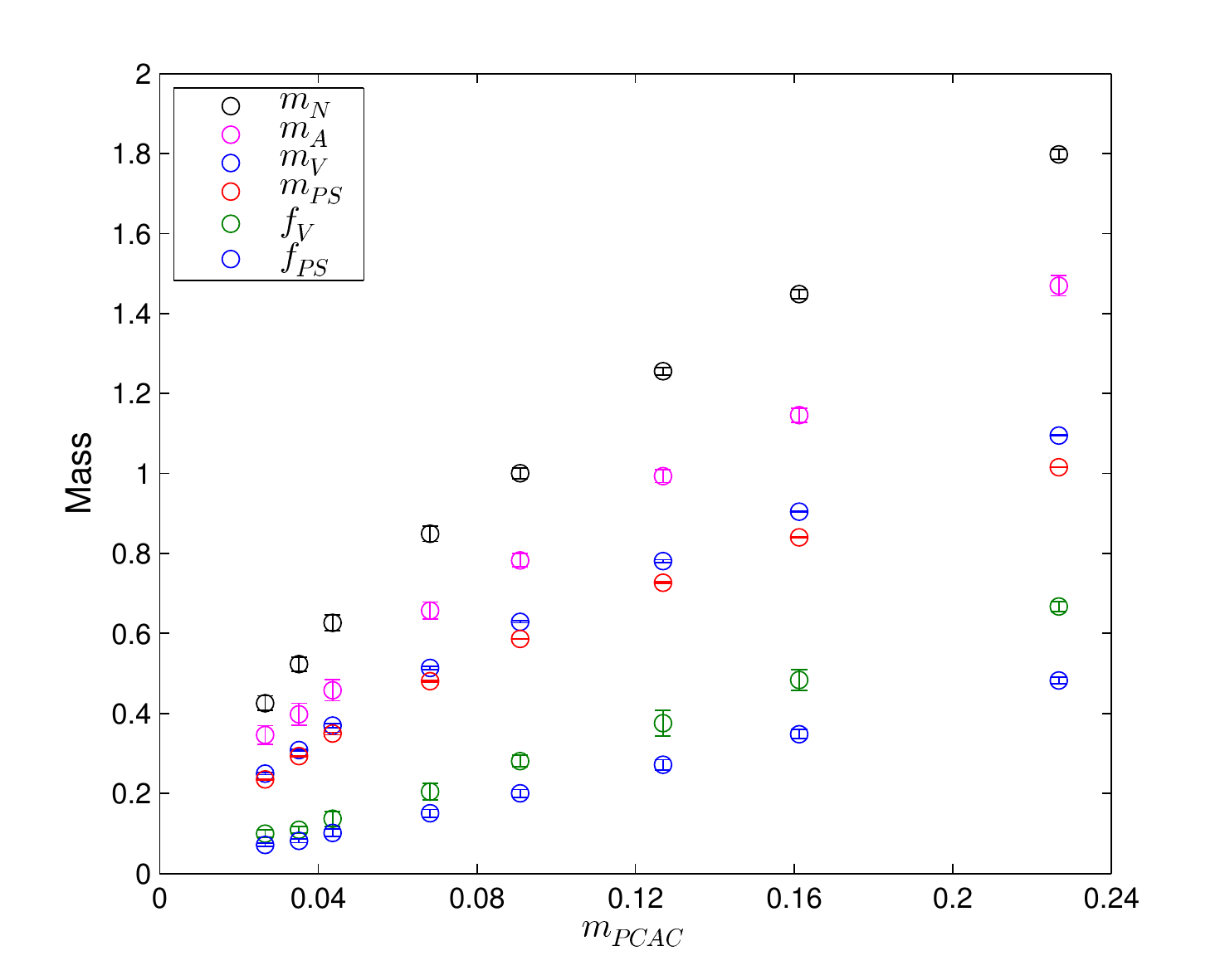}
 \includegraphics[scale=0.50]{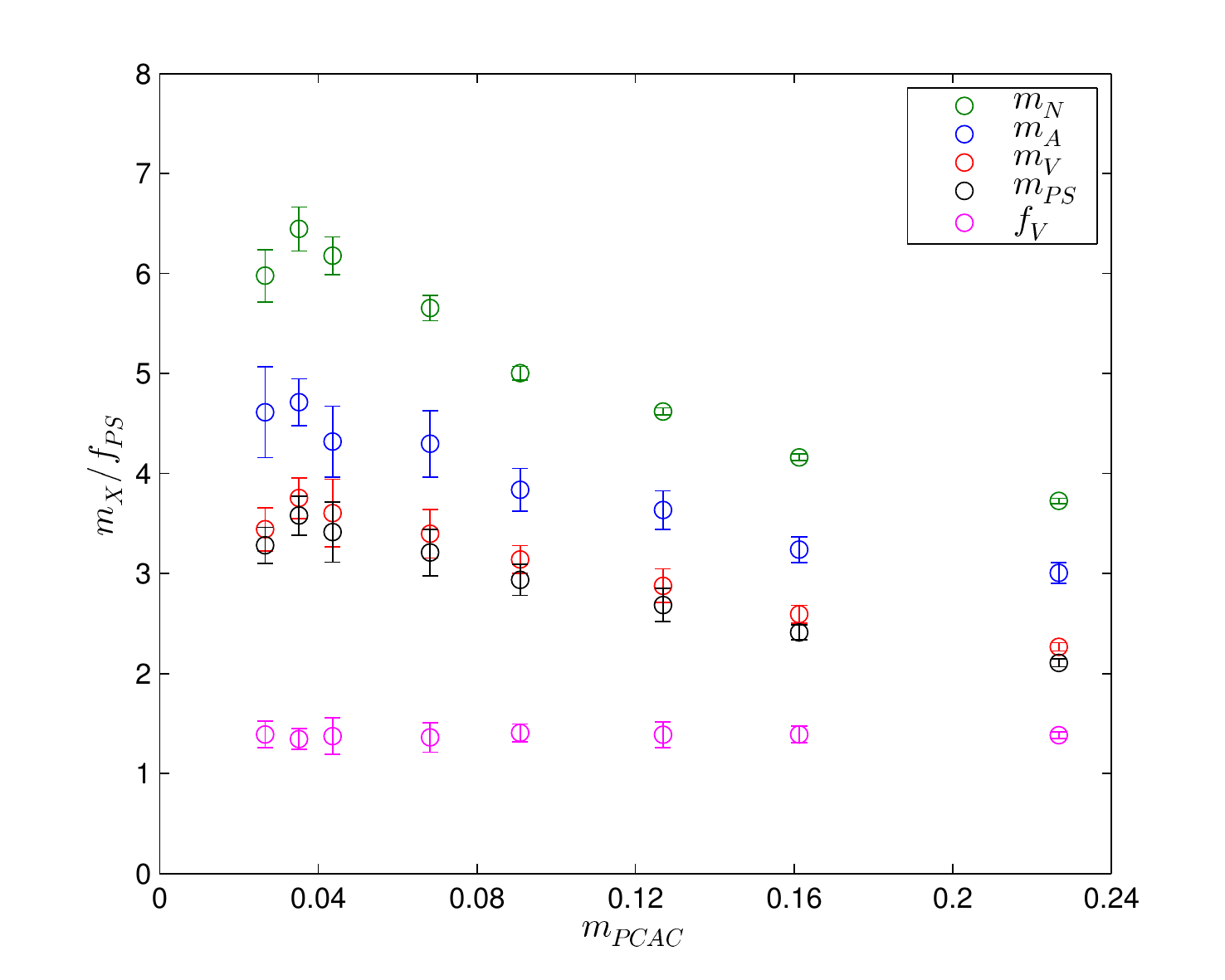}
\end{center}
\vspace{-5mm}
\caption{\textit{Left}: Overview of the spectral quantities considered for the analysis in section~\ref{sect:spectrum}. \textit{Right}: The same quantities normalized to $f_{PS}$.}
\label{fig:spectrum}
\end{figure}

\subsection{Finite volume effects}\label{sect:fvolume}
\begin{figure}
 \centering
 \includegraphics[scale=0.54]{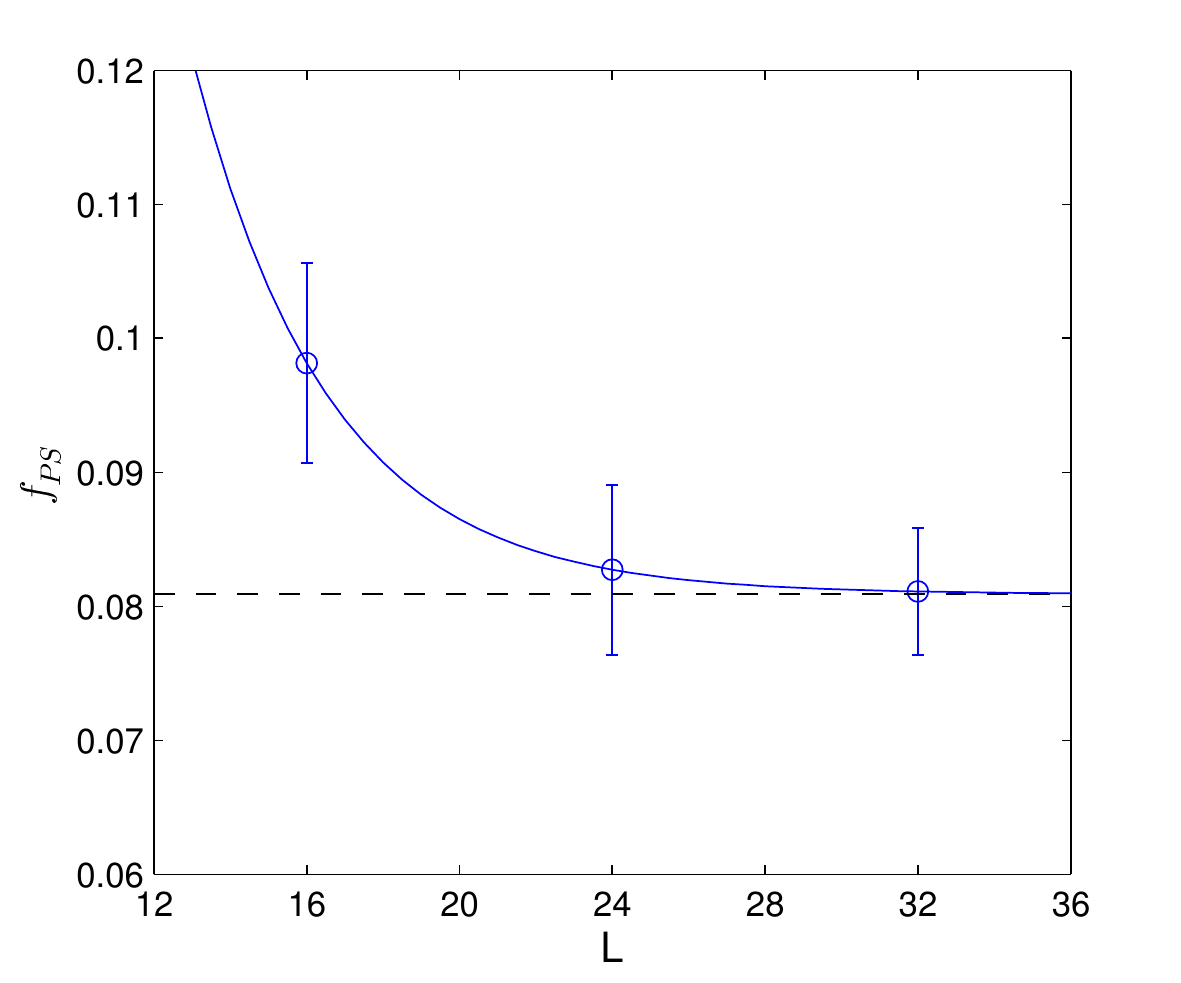}
 \includegraphics[scale=0.54]{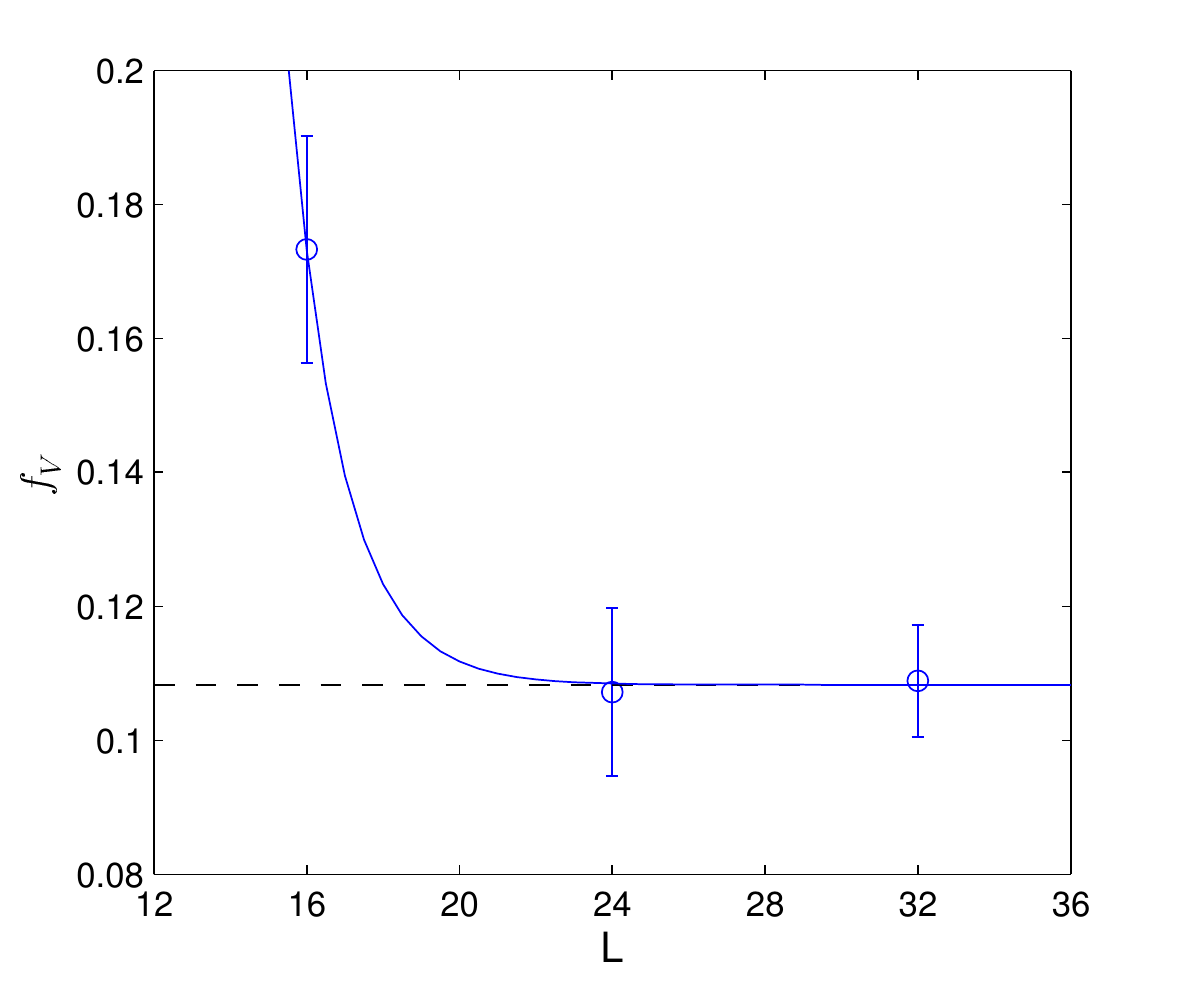}
 \includegraphics[scale=0.54]{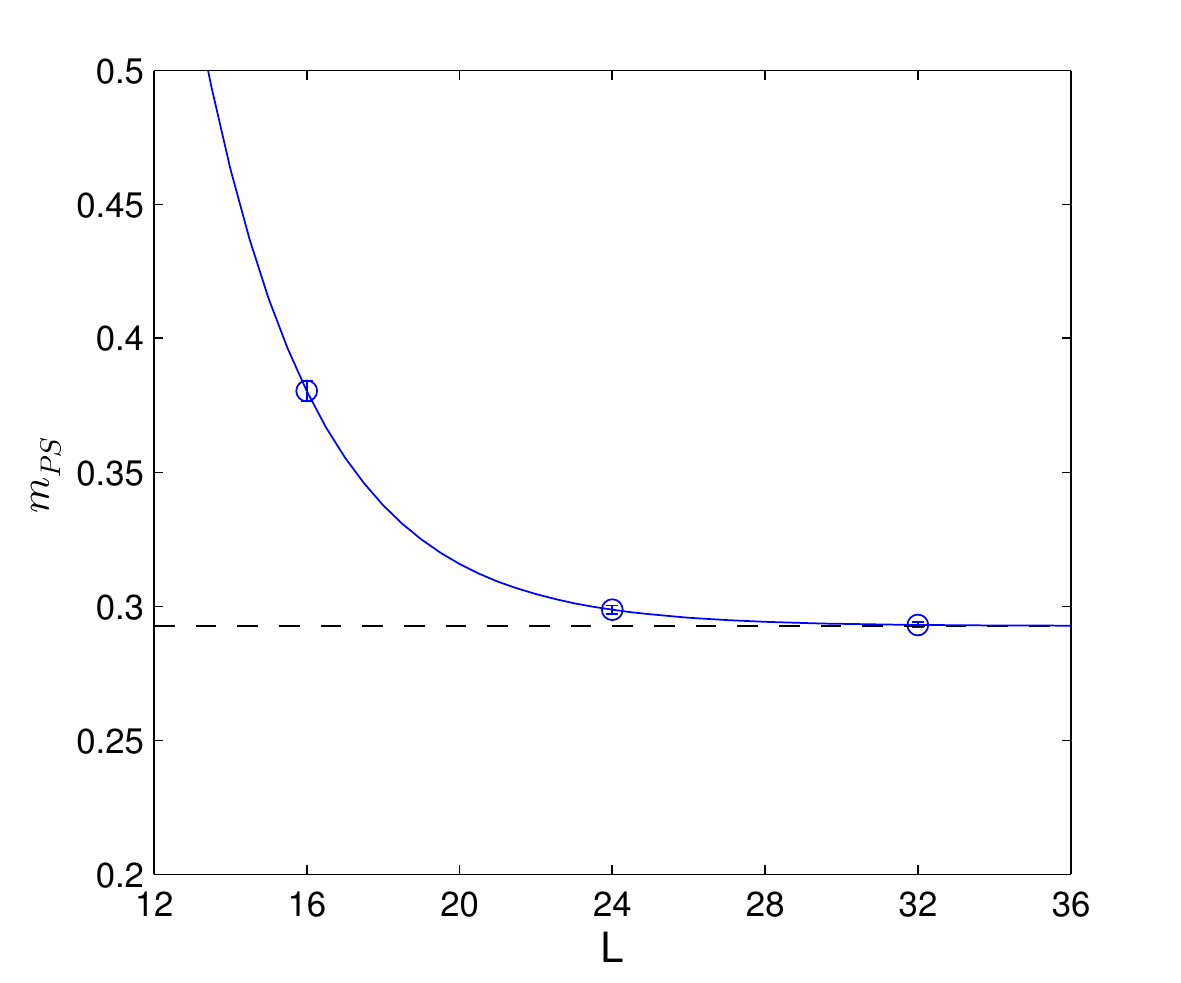}
 \includegraphics[scale=0.54]{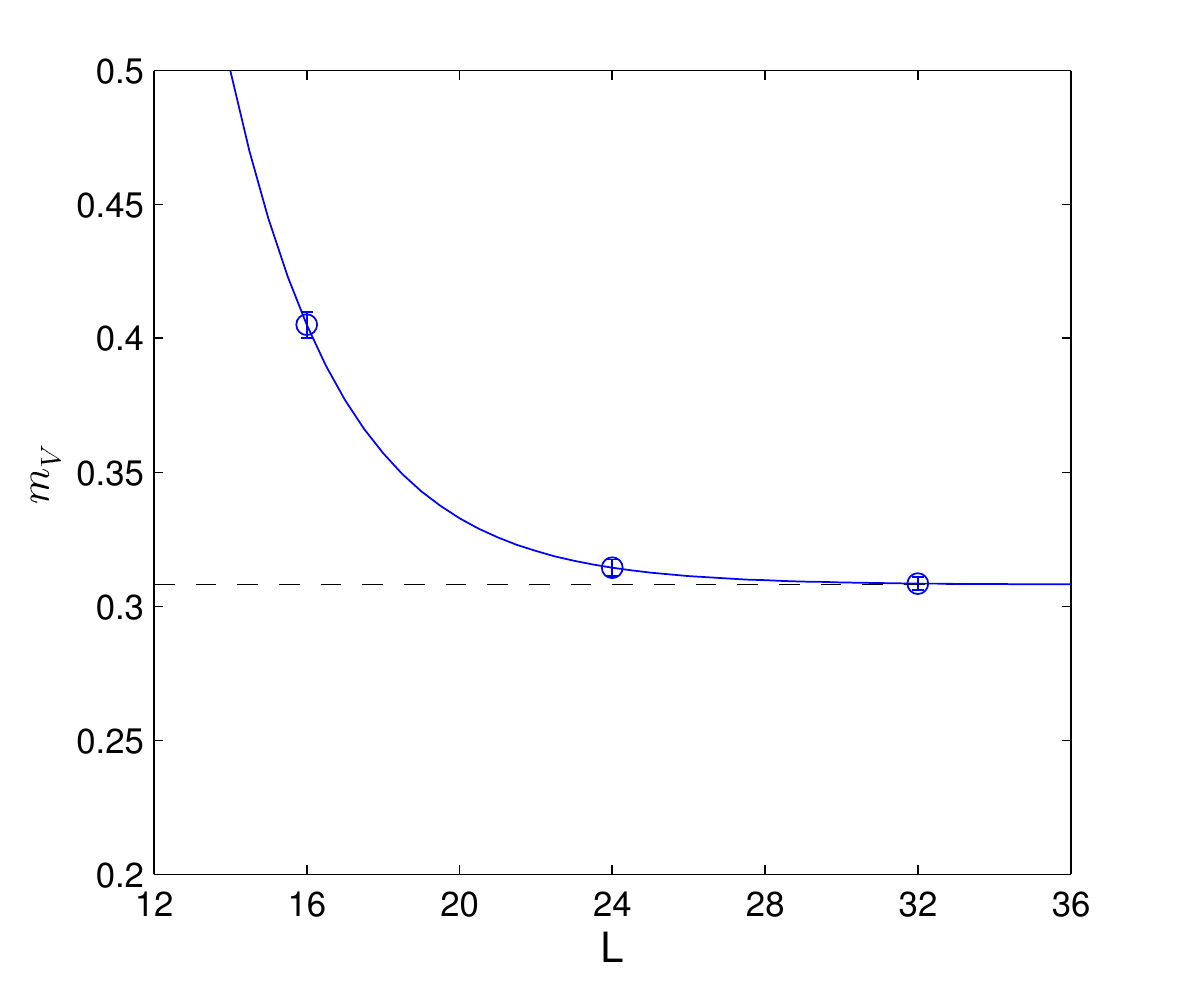}
 \includegraphics[scale=0.54]{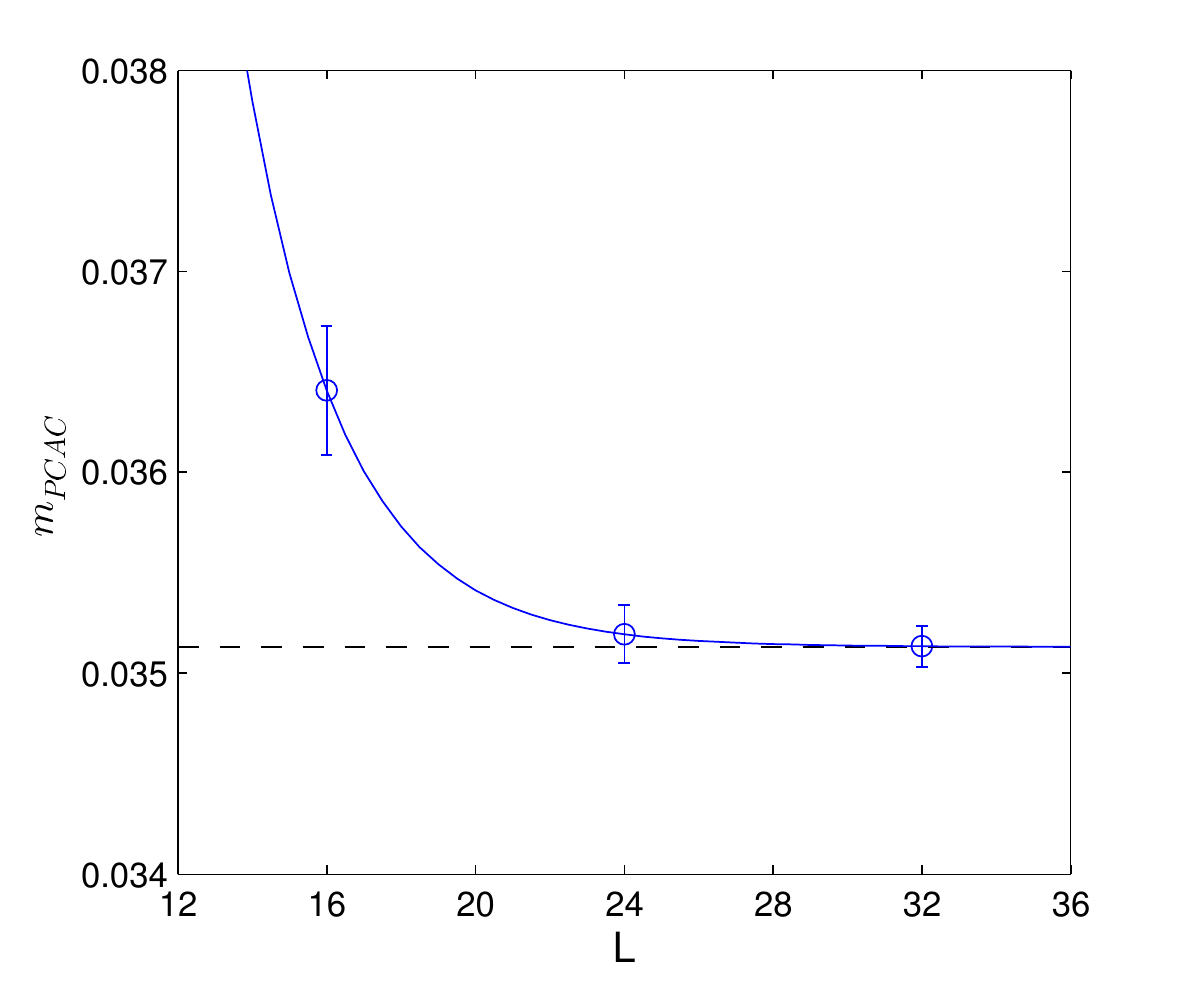}
 \caption{Measure of finite volume effects at $\beta=5.4$ on the second lightest point available for this study. All finite volume effects are controllably small on the largest lattice used $32^3\times 48$.}
 \label{fig:finitev}
\end{figure}

We have studied the finite volume effects for the observables considered in this work. We show in Fig.~\ref{fig:finitev} the finite volume effects on $m_{PCAC}$, $f_{PS}$, $m_{PS}$, $f_{V}$ and $m_V$ for the second lightest point considered in this section, corresponding to a pseudoscalar meson mass of $m_{PS}=0.293(1)$. To quantify the finite volume effects we use three different spatial volumes with $L=\{16,24,32\}$ and the temporal extent fixed to $T=48$ in all cases. To parametrize the finite volume corrections we use an exponential function of the form
\begin{equation}
 m(L) = m_\infty + a\exp(-bL),
\end{equation}
with $\{m_\infty,a,b\}$ as free parameters, which fits out data very well. On the largest volume, the finite volume effects are negligible and in all cases much smaller than the statistical errors.
We also note that, in the case of the pseudoscalar decay constant, the finite volume corrections have an opposite sign when compared to the expected behavior of a chirally broken theory. From these results we conclude that the systematic errors stemming from the finite simulation volume are under control for the spectral quantities considered in this section.

\subsection{Spectrum}\label{sect:spectrum}
It is in principle possible to distinguish between the two scenarios by studying the particle spectrum in the chiral limit. The most dramatic difference between the two scenarios is that in the IR conformal case all meson masses and decay constants will vanish in the chiral limit, while in the more familiar chirally symmetric broken case only the mass of the Nambu-Goldstone bosons will become zero. 
In both cases analytical predictions are available for the behavior of meson masses and decay constants, which will be reported below. Because both predictions are perturbative expansions around the quark mass, they are only expected to work sufficiently close to the chiral limit.

Here we analyse the spectral quantities $f_{PS}$, $m_{PS}$, $f_{V}$, $m_V$, $m_A$ and $m_N$ as a function of $m_{PCAC}$.
We show in Fig.~\ref{fig:spectrum} an overview of all the quantities considered here (left panel) and the same quantities normalized to $f_{PS}$.
The numerical values obtained in our simulations are reported in Table~\ref{tab:b54} and \ref{tab:b54ratio} in Appendix~\ref{sec:appA}.

As noted previously in Section~\ref{sect:pd-spec}, we observe that all the spectral quantities considered have a strong dependence on the quark mass and the ratios to $f_{PS}$ seem to converge to a finite value in the chiral limit. 
In particular this is also the case for the ratio $m_{PS}/f_{PS}$ which slightly increases towards the chiral limit. 
This kind of behavior has been observed in all previous studies of conformal and near conformal models, such as e.g. the SU(3) model with $n_f=8$ fundamental Dirac fermions~\cite{Aoki:2016wnc,Appelquist:2016viq} and the sextet model considered here~\cite{Fodor:2016pls}.   

If such a behavior would persist for arbitrarily small quark masses, the theory would be IR conformal. On the other hand if the mass of $m_{PS}$ vanishes while all the other spectral quantities remain finite, the theory would be chirally broken. As all the quantities show a strong dependence on the quark mass, a careful analysis is required to distinguish between the two cases. 

As we will show below, similarly to all other numerical studies of (near) IR conformal models, the expected leading order behavior for both scenarios is only attained for very small quark masses. Therefore corrections to the leading order behavior must be included in the analysis. 

We now study how well each of the two hypothesis for the chiral behavior of the model fit our data.

\subsection{Test of conformal scaling}
We first consider the case of an IR conformal model. 
In an IR conformal theory the RG equations can be used to show that all masses $M_x$ and decay constants $F_x$ scale in the same way as a function of the quark mass \cite{DelDebbio:2010hu,DelDebbio:2010jy,DelDebbio:2010ze}, in a neighbourhood around the IR fixed point. The scaling behaviour is described by a hyperscaling relation of the form\footnote{For the decay constants we only consider $f_{PS}$ and $f_V$ which have the same scaling exponents as the masses.}
\begin{align}
\begin{split}
 M_x &= A_x m^{1/(1+\gamma)} + \tilde{A}_x m^{\omega} \,,\\
 F_x &= B_x m^{1/(1+\gamma)} + \tilde{B}_x m^{\omega}\,,
\end{split}
\label{eq:conformal_fit}
\end{align}
where $m$ is the quark mass, $x$ is a label for a specific channel, and $\gamma$ is the the anomalous dimension of the mass at the IR fixed point which is related to the universal leading scaling exponent. 
The second term in each expression is the dominant sub-leading correction to scaling, where the dominant subleading exponent $\omega$ is also universal and bigger than $1/(1+\gamma)$.
In particular, this scaling behaviour predicts that the ratio of any two observables is constant up to higher order corrections and that all particles are massless in the chiral limit.

\begin{figure}[H]
\begin{center}
 \vspace{-3mm}
 \includegraphics[width=0.48\textwidth]{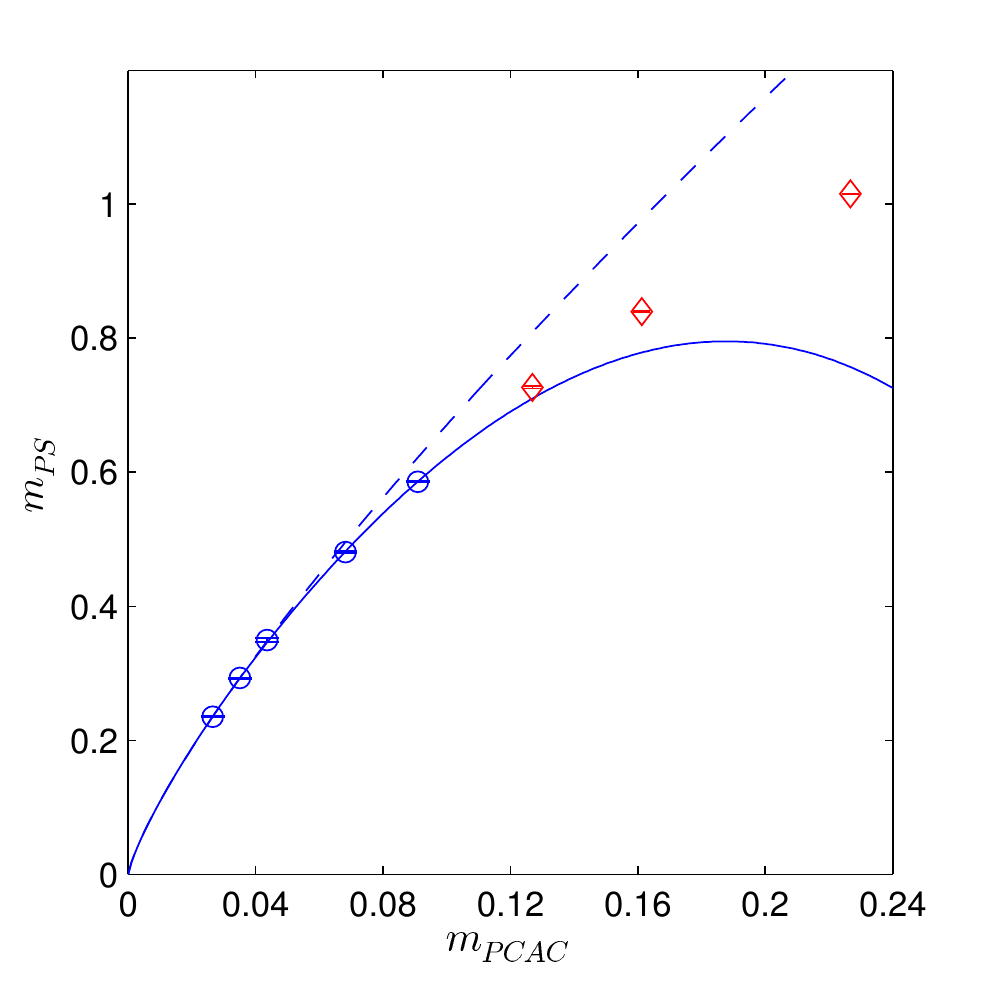}
 \includegraphics[width=0.48\textwidth]{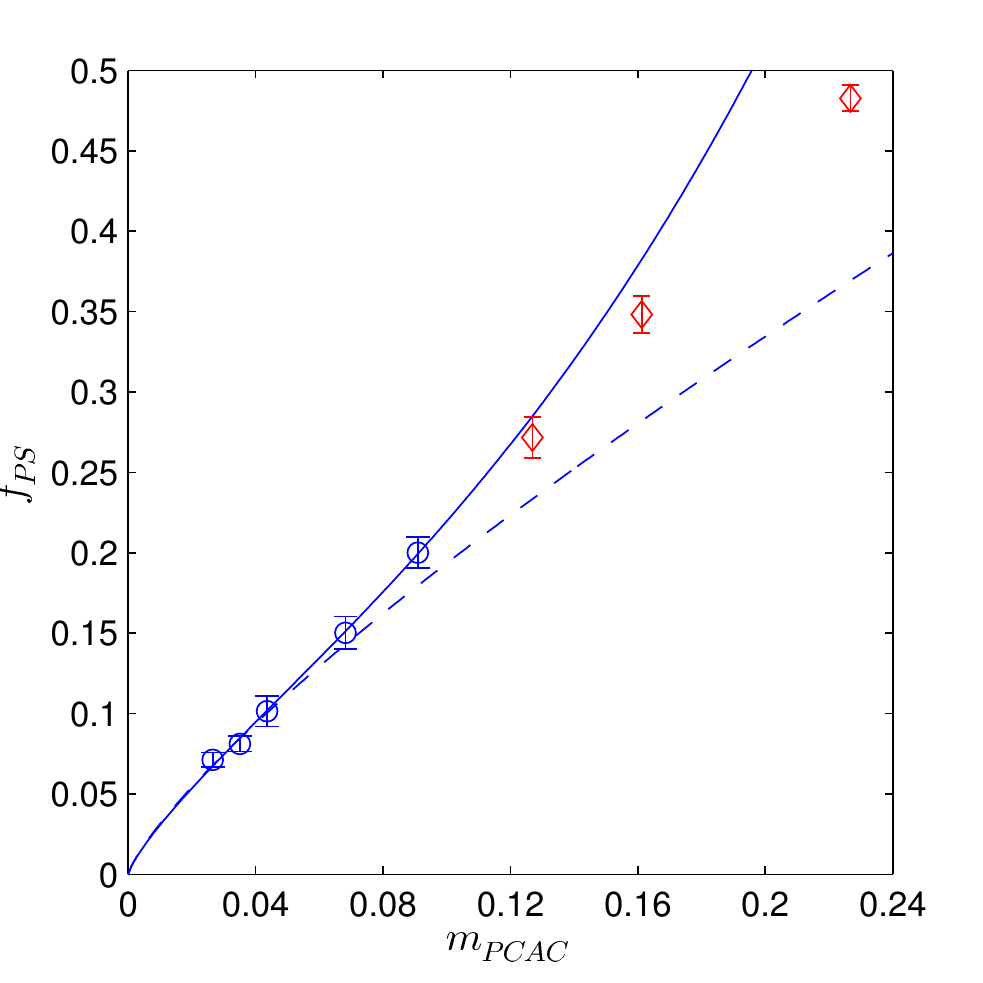} \\
 \vspace{-3mm}
 \includegraphics[width=0.48\textwidth]{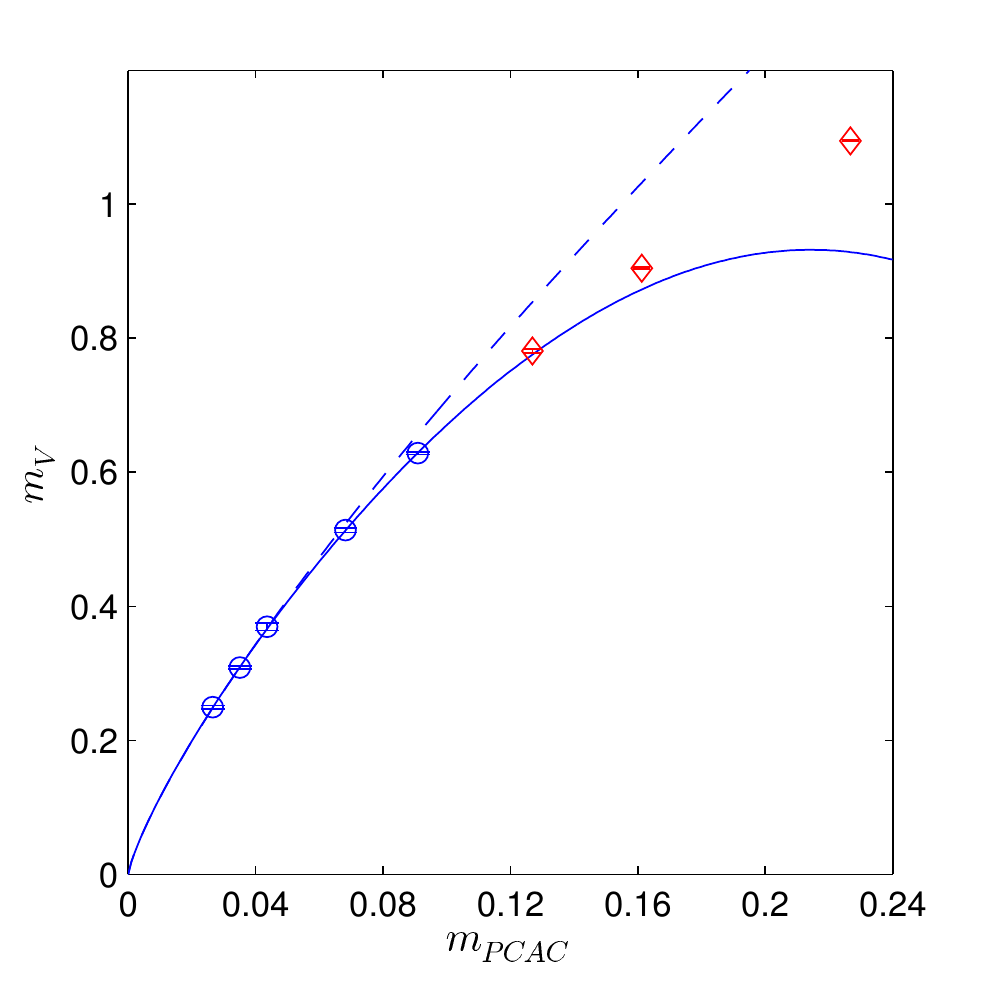}
 \includegraphics[width=0.48\textwidth]{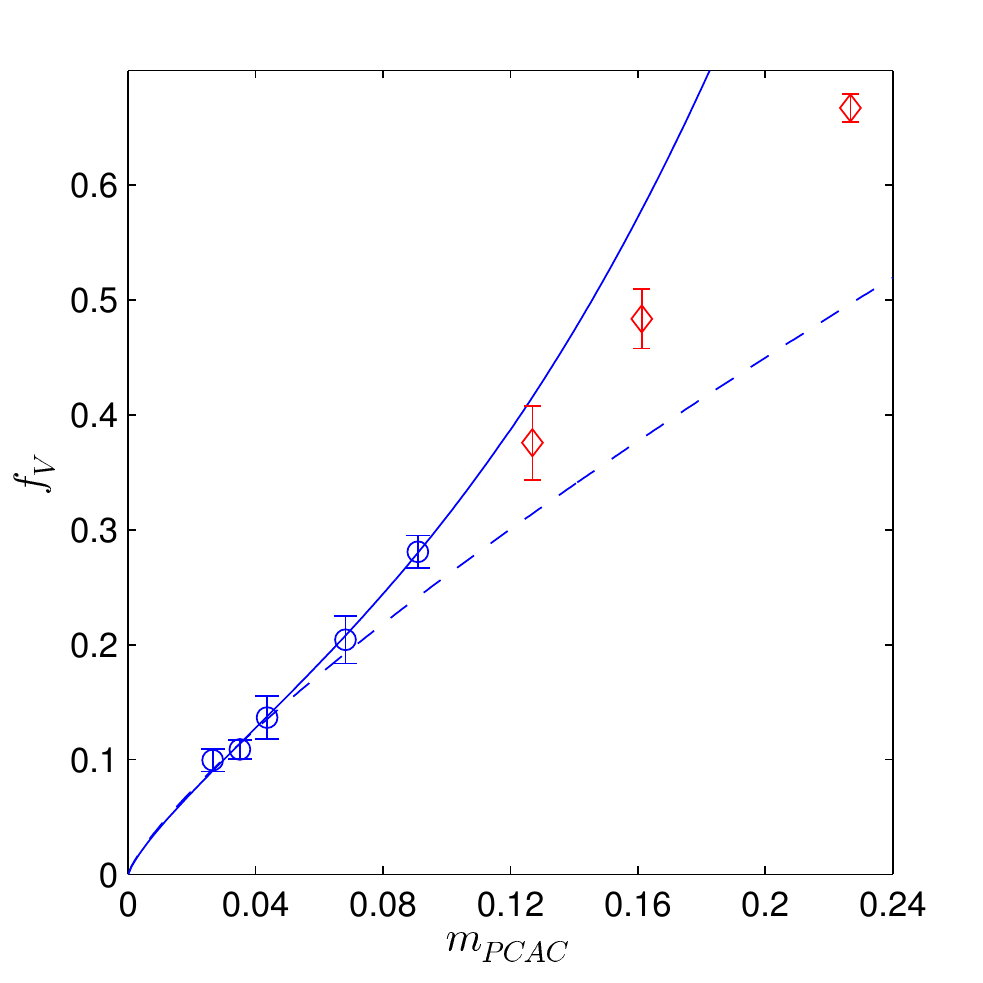} \\
 \vspace{-3mm}
 \includegraphics[width=0.48\textwidth]{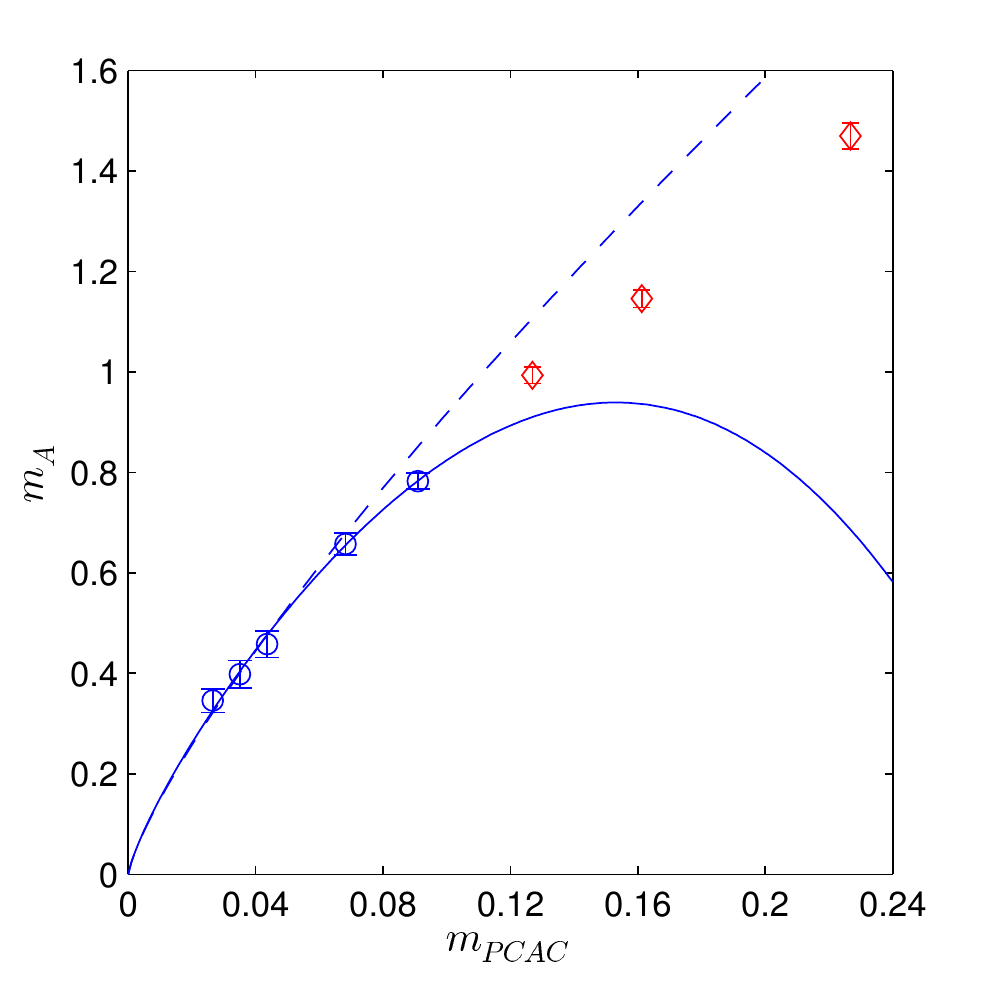}
 \includegraphics[width=0.48\textwidth]{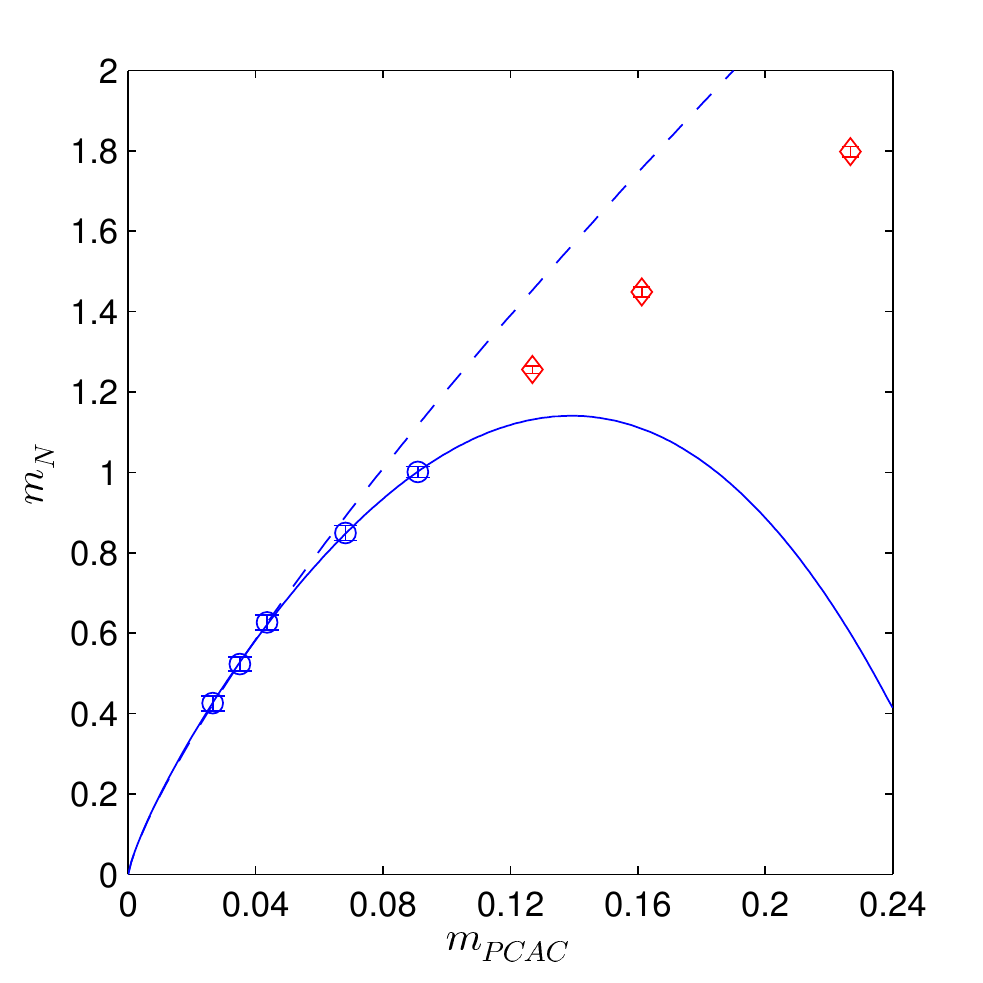} \\
 \vspace{-3mm}
\end{center}
\caption{Combined conformal fits for $\beta=5.4$ with and without subleading corrections. 
The dashed line is the leading order conformal fit to the three lightest points in all channels. The solid line is an independent fit with subleading scaling corrections, to the five lightest points in all channels. The data used, reported in Table~\ref{tab:b54}, is from simulations at $\beta=5.4$ and infinite volume.}
\label{fig:conformal}
\end{figure}

We show in Fig.~\ref{fig:conformal} the result of the combined fit to all the channels available in this study. The fit to Eq.~\eqref{eq:conformal_fit} (solid line in the figure) describes the data well over a range of quark masses up to $m_{PCAC}\sim 0.10$ with a $\chi^2/\mathrm{dof}=7.04/16=0.44$.
The fitted value for the anomalous dimension of the mass is $\gamma=0.25(3)$ and the subleading exponent is $\omega=2.71(76)$.
We note that in this case subleading corrections are crucial to obtain a good fit, mainly because they give a large contribution to the decay constants. 
We also show the result of the best fit to the leading order behavior (dashed line in the figure), where we reduce the range of masses included in the fit to $m_{PCAC}<0.05$. 
Over this reduced mass range the fit to leading order behavior also provides a good description of our data with $\chi^2/\mathrm{dof}=4.62/11=0.42$.
The best fit value for the mass anomalous dimension extracted from this fit is $\gamma=0.27(3)$ which is in very good agreement with the value obtained considering corrections to scaling.
This is an indication that subleading corrections are indeed a small correction at small quark masses, which we also verified explicitly.
We have also investigated the stability of the fit to Eq.~\eqref{eq:conformal_fit} when changing the range of masses included in the fit. 
We found that including heavier points makes the subleading corrections dominant even at small quark masses, which we consider unacceptable.
On the other hand, if we reduce the range of fitted masses, the fit remains stable.

\subsection{Test of chiral symmetry breaking}
We now turn to the case of spontaneous chiral symmetry breaking.
The CCWZ formalism can be used to write down an effective low energy Lagrangian in the chiral limit, as done in chiral perturbation theory. 
Because the sextet representation is complex, the pattern of chiral symmetry breaking is $SU(2)_L\times SU(2)_R\to SU(2)_V$, i.e.~the same as in two-flavor QCD. 
The quark mass dependence of the pion mass and the pion decay constant has been calculated to next-to-next-to-leading (2-loop) order in \cite{Bijnens:2009qm}:
\begin{align}
\begin{split}
 M_{\pi}^2 &= M^2 \left[ 1 + \frac{M^2}{F^2}(a_ML+b_M) + \frac{M^4}{F^4}(c_ML^2+d_ML+e_M) \right]\,,\\
 F_{\pi} &= F \left[ 1 + \frac{M^2}{F^2}(a_FL+b_F) + \frac{M^4}{F^4}(c_FL^2+d_FL+e_F)\right] \,.
\end{split}\label{eq:cpt}
\end{align}
Here $M^2=2Bm$ is the leading order pion mass from the GMOR relation, $F$ is the leading order pion decay constant, and $L$ is short-hand notation for the chiral logs:
\begin{equation}
 L = \frac{1}{16\pi^2}\log\left(\frac{M^2}{\mu^2}\right)\,.
\end{equation}
The coefficients $\{a_M,a_F,c_M,c_F\}$ are known in the continuum limit and have values
\begin{equation}
 a_M = \frac{1}{2},\qquad
 a_F = -1,\qquad
 c_M = \frac{17}{8},\qquad
 c_F = -\frac{5}{4},\label{eq:coef}
\end{equation}
but the remaining coefficients are combinations of unknown low-energy constants.
On the lattice, the coefficients $\{a_M,a_F,c_M,c_F\}$ will receive $\mathcal{O}(a)$ corrections.

On a qualitative level, chiral perturbation theory predicts massless pions in the chiral limit (they are the Goldstone bosons) but with a non-zero decay constant. The remaining particles in the spectrum are expected to have a finite mass.
The chiral perturbative expansion is expected to work in the regime where the pion is the lightest state in the model and it is expected to converge in the limit of small expansion parameter $M^2/(4\pi F)^2\ll 1$.

\begin{figure}[H]
\begin{center}
 \vspace{-3mm}
 \includegraphics[width=0.48\textwidth]{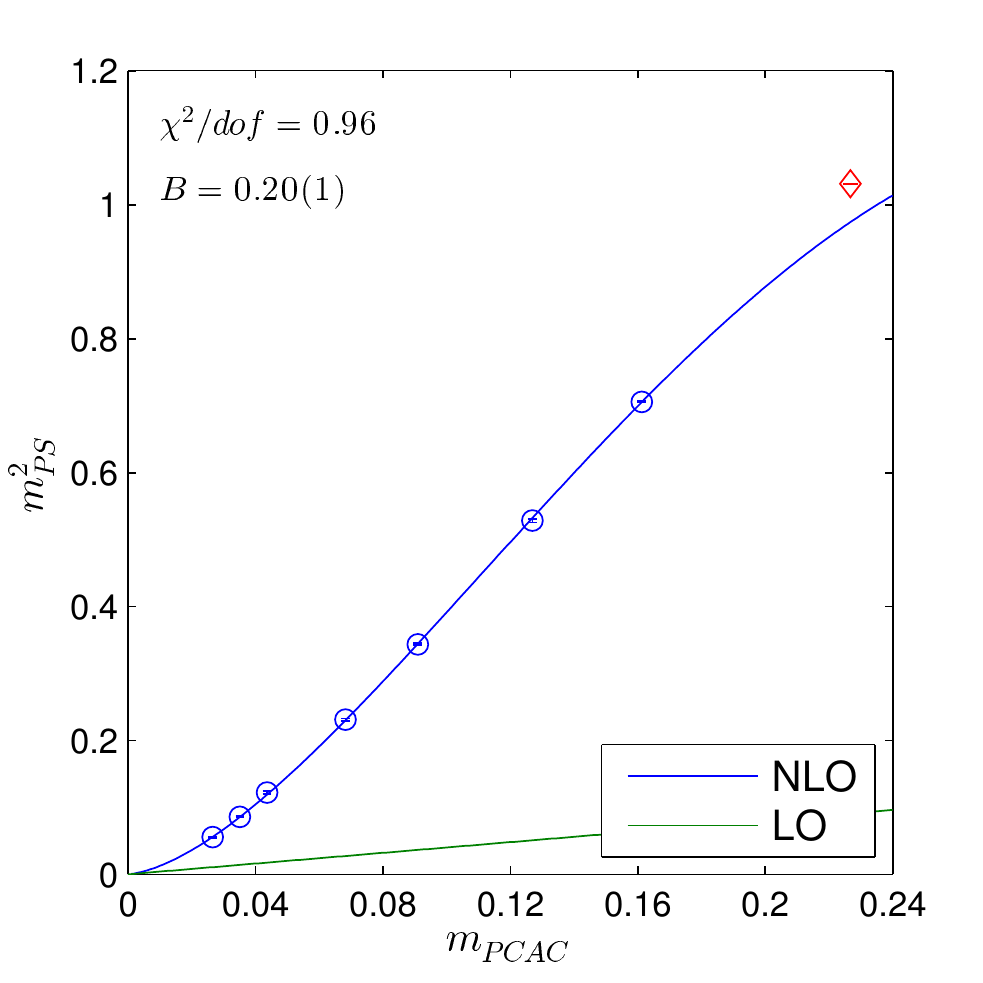}
 \includegraphics[width=0.48\textwidth]{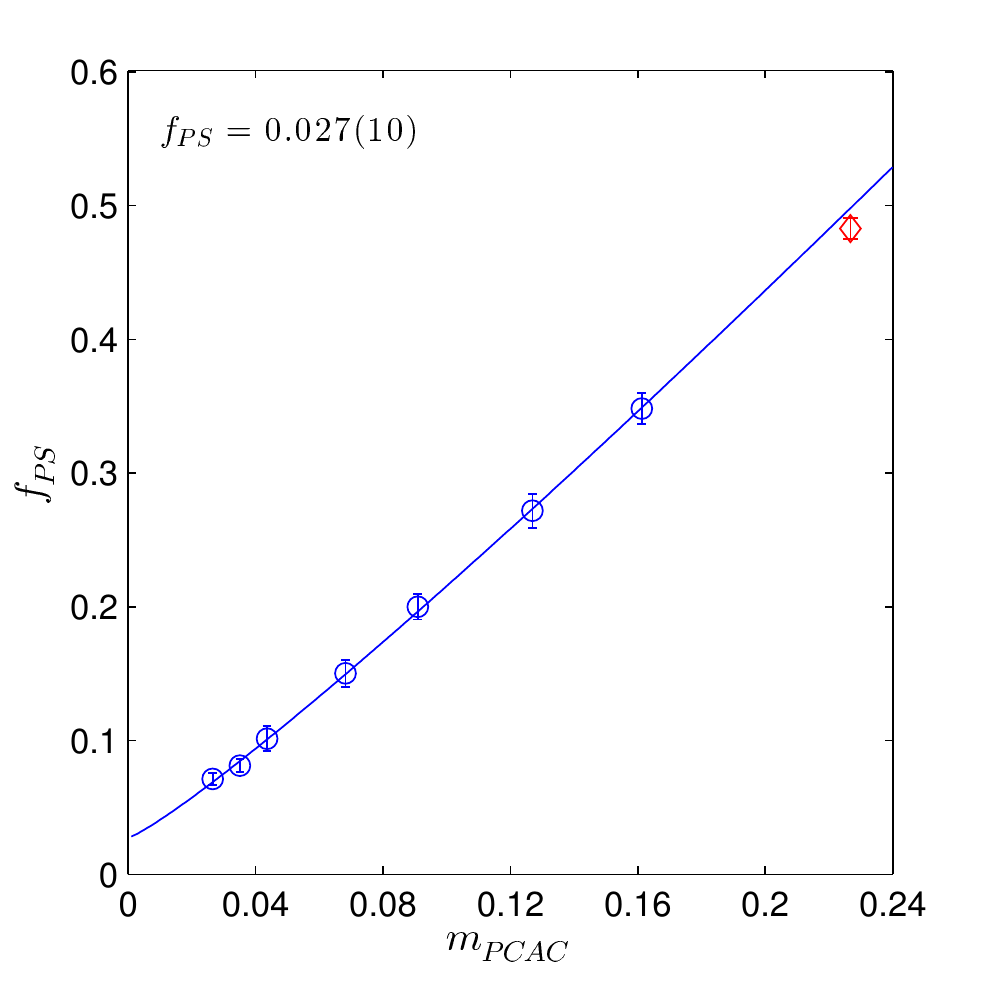} \\
 \vspace{-3mm}
 \includegraphics[width=0.48\textwidth]{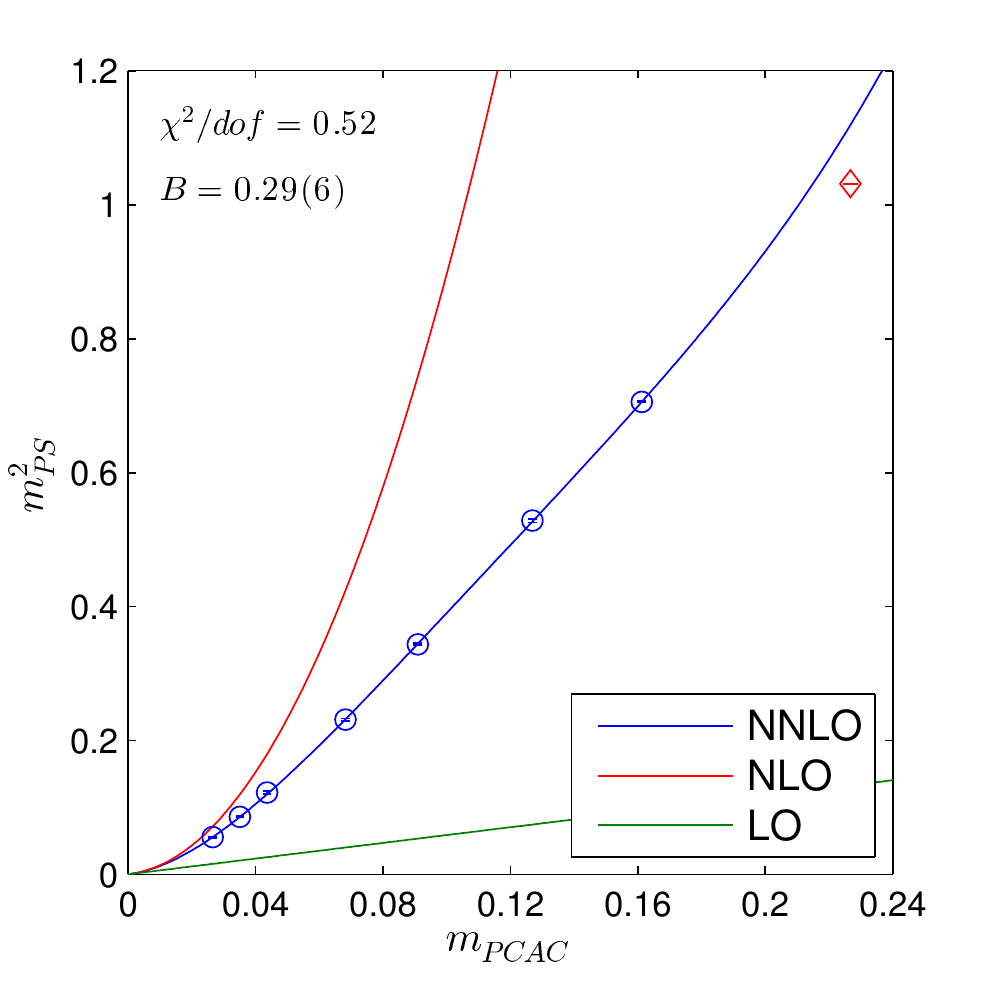}
 \includegraphics[width=0.48\textwidth]{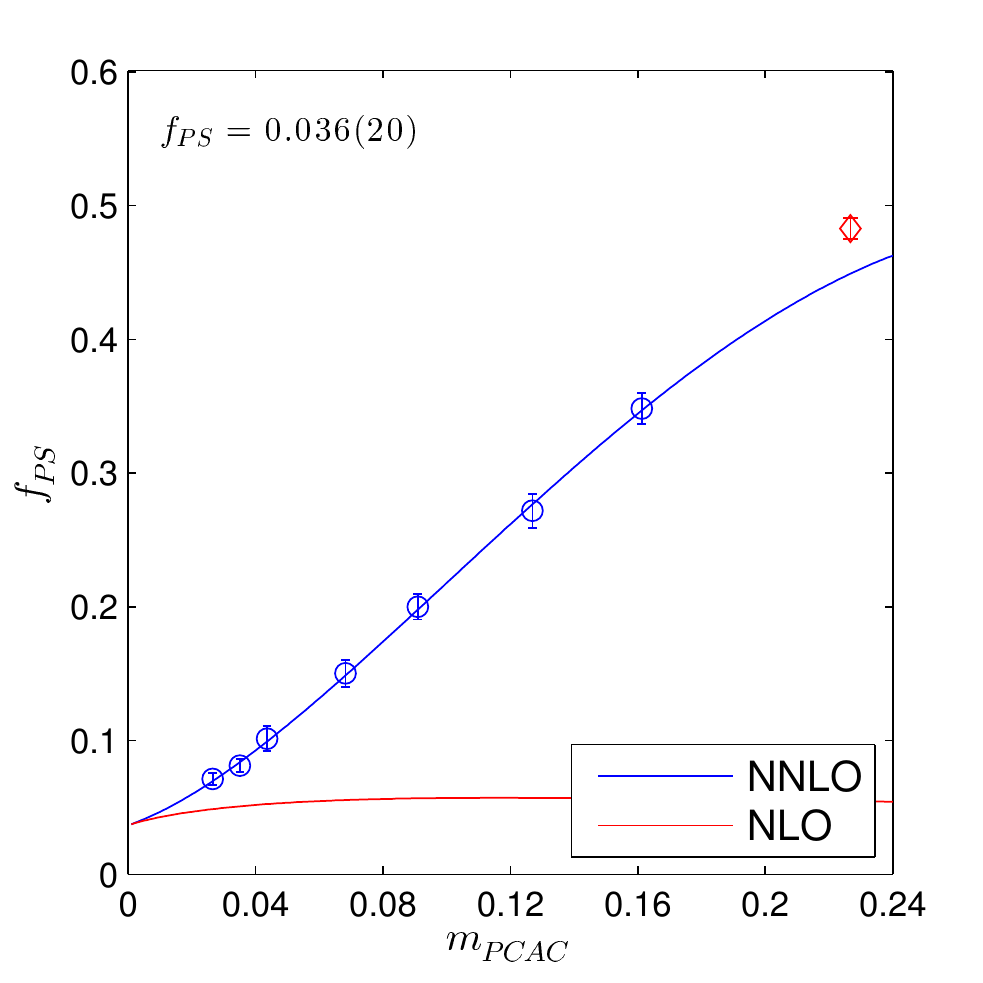} \\
 \vspace{-3mm}
 \includegraphics[width=0.48\textwidth]{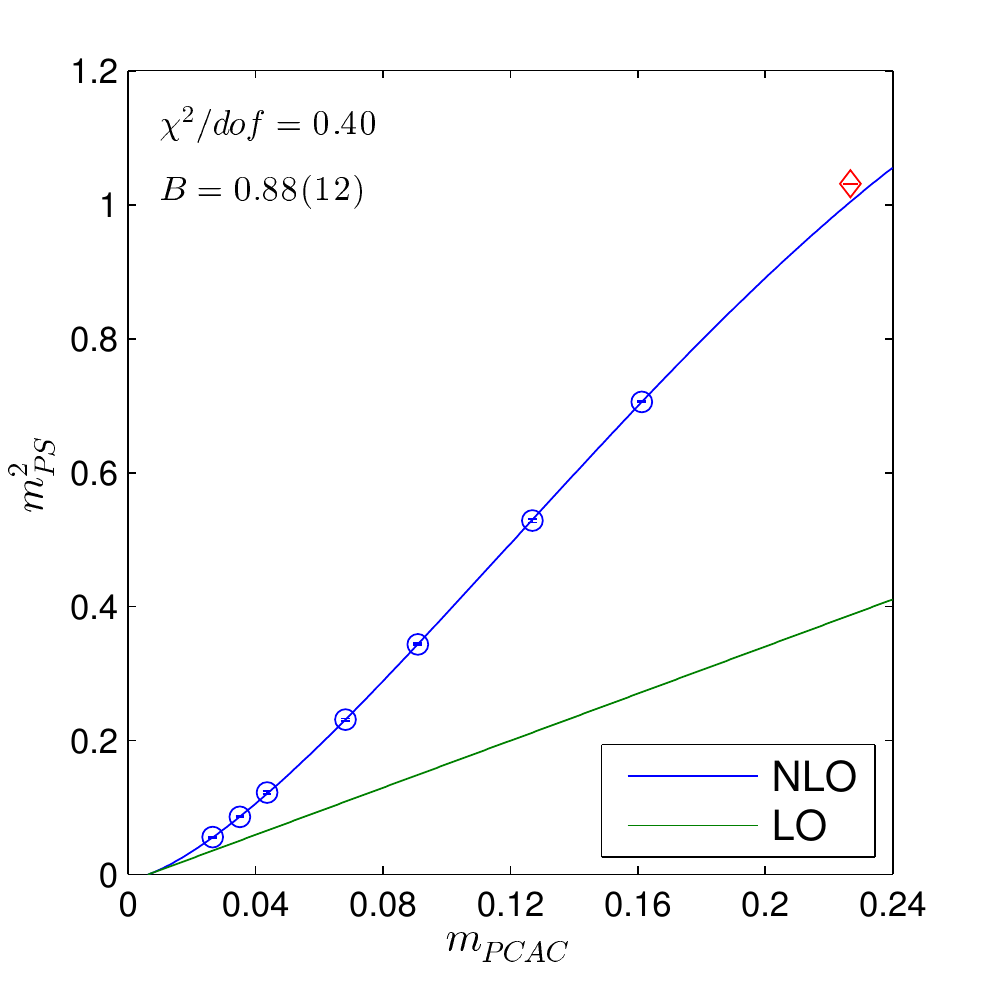}
 \includegraphics[width=0.48\textwidth]{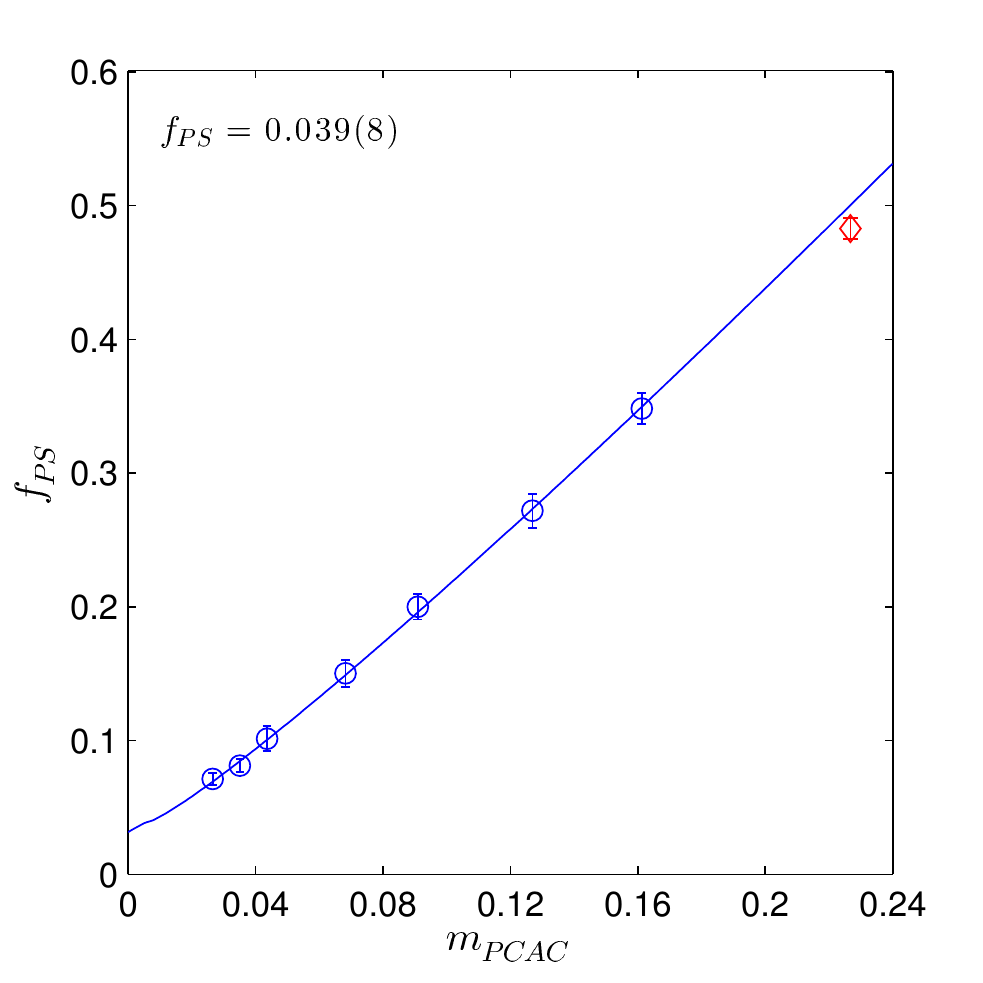} \\
\end{center}
\caption{Chiral fits for $\beta=5.4$. \textit{Top}: Result at NLO order with free coefficients for the log terms. \textit{Middle}: Result at NNLO order with continuum logs. \textit{Bottom}: NLO fit with free coffiecients and with an additional $\mathcal{O}(a^2)$ shift $M^2=2Bm+\delta$.}
\label{fig:chpt_nlo}
\end{figure}

We stress that the applicability of chiral perturbation theory to our data is questionable since over the entire range of quark masses explored in our study the mass of the vector meson is almost the same as the "pion". Additionally, in all other studies of (near) IR conformal models, such as the ones previously cited, it has been shown that a light scalar resonance, analogous to the $f_0(500)$ in QCD, is present, which can be as light (or even lighter) than the pion.

Despite these problems, we try to use Eq.~\eqref{eq:cpt} to fit our numerical results. We find that the expected next-to-leading order behavior with the coefficients $a_M$ and $a_F$ fixed to their continnum values does not fit the numerical data. 

Instead we perform two other fits which provide a good description of our results: a next-to-leading order fit with $a_M$ and $a_F$ as free coefficients; and a next-to-next-to-leading order fit with the coefficients in Eq.~\eqref{eq:coef} fixed.
We show in Fig.~\ref{fig:chpt_nlo} the resulting best fit for both the next-to-leading (top row) and the next-to-next-to-leading (middle row) formula. In both cases the same mass range $m_{PCAC}\le 0.17$ for the fit is used.
In the NLO case, the fit has a $\chi^2/\mathrm{dof}=5.76/6=0.96$ and we find $F=0.027(10)$ and $B=0.20(1)$.
For the NNLO case we obtain $\chi^2/\mathrm{dof}=2.08/4=0.52$ with $F=0.036(20)$ and $B=0.29(6)$.
The values obtained from the two fits for $F$ are in good agreement, while, in the case of $B$, the poorer agreement reflects the very large uncertainty on the leading order behavior of the data.

By changing the fitted mass range, we find that the values for $F$ are relatively stable around $F\approx 0.03$ with large errors, while $B$ seems to decrease, also significantly, if we reduce the mass range to smaller quark masses.

Finally we tried a modified version of Eq.~\eqref{eq:cpt} in which we introduce an $\mathcal{O}(a^2)$ shift in the leading order pion mass squared: $M^2=2Bm+\delta$. 
This functional form is inspired by Wilson chiral perturbation theory (WChPT). We show in Fig.~\ref{fig:chpt_nlo} (bottom row) the result of the fit which describes our data well. Also in this case we obtain a value of $F=0.039(8)$ which is compatible with the values above, while the best fit value of $B$ is, not surprisingly, substantially increased to $B=0.88(12)$.

In Fig.~\ref{fig:chpt_nlo} we also show the size of the different loop order terms to $m_{PS}$ and $f_{PS}$. It is clear, given the large size of higher order corrections, that the perturbative series is not convergent in the fitted mass range. 
Even for our best fit, inspired by WChPT, the NLO terms are clearly dominating the LO terms over the whole range of mass explored. 
Therefore, even if one believes that Eq.~\eqref{eq:cpt} give a correct description of the system, our numerical data are not within its range of applicability.

In view of these results, althought it is possible to use a functional form inspired by chiral perturbation theory to describe our numerical data, we conclude that little physical meaning can be extracted from this analysis.
If the model is chirally broken, substantially lighter quark masses are needed to make contact with ChPT.


\section{Conclusions}
We presented a detailed study of the SU(3) ``sextet" model with two Dirac fermions in the two-index symmetric representation of the gauge group. 
The main phenomenological motivation to consider such a model stems from the possibility that it is a realistic candidate for a Walking Technicolor model if it sits just below the sill of the conformal window. 
The precise location of the conformal window is also an interesting theoretical question, which many groups have investigated on the lattice. 
For the sextet model this issue is still controversial \cite{Fodor:2012ty, Fodor:2014pqa, Fodor:2016wal, Fodor:2016pls,Pica:2017gcb}.

In this paper we studied the infrared properties of the model by focusing on the physical spectrum.  
The most precise results for the spectrum of the model have, so far, been obtained with the use of the staggered fermion discretization. In this work we chose the Wilson discretization for a comparison.

We mapped out the phase diagram of the lattice model and located several distinct regions in the parameter space, see Fig.~\ref{fig:phase_diagram}, which show qualitatively very different behaviors. 
In particular, we locate a strong coupling ``bulk" phase in which the model shows the qualitative features expected from spontaneous chiral symmetry breaking, such as a significant splitting between the would-be Goldstone boson mass and the vector meson mass. However in this strong coupling phase, vanishing pion masses cannot be reached because of a first order transition that occurs at small quark masses. 

At weak couplings this first order line disappears, so that one can identify a chiral line where the quark mass vanishes.  However in this weak coupling phase, we do not observe the qualitative features expected from a chirally broken model, e.g. the ratio between the pseudoscalar and vector meson masses remain close to one and constant over the entire range of masses explored in this study.

A similar abrupt change in behavior is seen in the scale setting observables $t_0$ and $w_0$, which are almost insensitive to the quark mass in the strong coupling phase, but strongly depend on it in the weak coupling phase where they seem to diverge in the chiral limit. 

We have checked that these qualitative features are free of finite volume effects and they would imply that the model is IR conformal, if they persists to vanishingly small quark masses. We therefore studied in more detail the chiral limit of the model in the weak coupling phase. We performed simulations at eight different quark masses and measured $f_{PS}$, $m_{PS}$, $f_{V}$, $m_V$, $m_A$ and $m_N$ and studied the behavior of these observables as a function of $m_{PCAC}$.
We compare our numerical data  to the predicted behavior for both the case of an IR conformal and a chirally broken model.

We find that it is possible to fit our data to the functional forms predicted in both cases.
However, in the case of IR conformality the best fit function is consistent with the theoretical expectations that subleading terms are small compared to the leading order scaling behavior. In particular, a fit to the simple leading order scaling behavior near the chiral limit is consistent with the fit which includes scaling corrections on an larger mass range.
In contrast, for the case of the chiral symmetry breaking, although we used several functional forms inspired by chiral perturbation theory, which describe our numerical data well, in all cases higher order corrections to the leading behavior are very dominant over the whole range of masses explored. 
This implies that the use of ChPT is questionable and significantly smaller quark masses are needed to observe a possible breaking of chiral symmetry.
The validity of ChPT is also questioned by the presence of a ``light" vector resonance and possibly an even lighter scalar resonance, as shown by studies with staggered fermions.

We conclude that the simplest interpretation of our data is that the model is IR conformal. The possibility remains that the model is ``walking" and it will eventually show signs of chiral symmetry breaking at much lighter quark masses than the ones used here.

In the future we plan to use an improved Wilson fermion setup to repeat the numerical study in the light quark region. This will allow for the use of coarser lattices and larger physical volumes, which are the main limiting factor in exploring the chiral limit of ``walking" or IR conformal models.

\acknowledgments
This work was supported by the Danish National Research Foundation under grant number DNRF90 and by a Lundbeck Foundation Fellowship grant. The computing facilities were provided by the Danish Center for Scientific Computing and the DeIC National HPC Centre at the University of Southern Denmark.

\appendix
\section{Sextet representation}\label{sect:sextet}
Fermions in the two-index symmetric representation can be written as
\begin{equation}
 \psi^{cd} = (\tilde{\psi}^ae^a)^{cd},
\end{equation}
where $\{\tilde{\psi}^a\}$ are the six Grassmann-valued degrees of freedom and $\{e^a\}$ is the orthonormal basis for all real and symmetric $3\times3$ matrices. For simplicity we only write the color indices, since the spinor indices are independent of the representation. Because this a symmetric representation we necessarily have  $\psi^{ab}=\psi^{ba}$ and for this reason the spinor field must transform as
\begin{align}
 \psi &\to U\psi U^T
\end{align}
where $U$ is an element of SU(3) in the two-index symmetric representation. An interesting feature of this model, is the existence of a baryon spectrum analogous to that of QCD. This can be seen by working out the color structure of a three-quark state, which contains a neutral color singlet.
\begin{equation*}
 \mathbf{6} \otimes \mathbf{6} \otimes \mathbf{6}
 = \mathbf{1} \oplus 2\times\mathbf{8} \oplus \mathbf{10} \oplus \overline{\mathbf{10}} \oplus 3\times\mathbf{27} \oplus \mathbf{28} \oplus 2\times\mathbf{35}.
\end{equation*}
The flavor structure for the baryons is equivalent to the flavor structure in QCD and it results in two doublets with mixed symmetry and a symmetric quadruplet.
\begin{equation*}
 \mathbf{2} \otimes \mathbf{2} \otimes \mathbf{2} = 2 \times \mathbf{2} \oplus \mathbf{4}
\end{equation*}
The quadruplet corresponds to the spin-3/2 baryons, the equivalents of the $\Delta$ baryons.
\begin{align}
\begin{split}
 \bra{\tfrac{3}{2},\tfrac{3}{2}}  &= uuu \\
 \bra{\tfrac{3}{2},\tfrac{1}{2}}  &= \frac{uud + udu + duu}{\sqrt{3}} \\
 \bra{\tfrac{3}{2},-\tfrac{1}{2}} &= \frac{ddu + dud + udd}{\sqrt{3}} \\
 \bra{\tfrac{3}{2},-\tfrac{3}{2}} &= ddd
\end{split}
\end{align}
The two doublets are equivalent representations of the spin-1/2 baryons. The first doublet is the mixed-antisymmetric (MA) representation
\begin{align}
\begin{split}
 \bra{\tfrac{1}{2},\tfrac{1}{2}} &= \frac{udd - dud}{\sqrt{2}}, \\
 \bra{\tfrac{1}{2},-\tfrac{1}{2}} &= \frac{udu - duu}{\sqrt{2}},
\end{split}
\end{align}
and the second doublet is the mixed-symmetric (MS) representation.
\begin{align}
\begin{split}
 \bra{\tfrac{1}{2},\tfrac{1}{2}} &= \frac{udd + dud - 2ddu}{\sqrt{6}}, \\
 \bra{\tfrac{1}{2},-\tfrac{1}{2}} &= -\frac{duu + udu - 2uud}{\sqrt{6}}.
\end{split}
\end{align}
With the group-theoretical knowledge of the model we can construct gauge invariant meson and baryon states. The meson states are constructed by contracting color indices with two Kronecker deltas.
\begin{align}
\begin{split}
 \delta^{ac}\delta^{bd}\bar{\psi}^{ab}\psi^{cd}
 &\to \delta^{ac}\delta^{bd}U^{*aa'}U^{*bb'}U^{cc'}U^{dd'}\bar{\psi}^{a'b'}\psi^{c'd'} \\
 &= (U^\dagger U)^{a'c'}(U^\dagger U)^{b'd'}\bar{\psi}^{a'b'}\psi^{c'd'} \\
 &= \delta^{a'c'}\delta^{b'd'}\bar{\psi}^{a'b'}\psi^{c'd'}
\end{split}
\end{align}
The baryon states can similarly be constructed by contracting the color indices with two Levi-Civita tensors.
\begin{align}
\begin{split}
 \epsilon^{ace}\epsilon^{bdf}\psi^{ab}\psi^{cd}\psi^{ef}
 &\to \epsilon^{ace}\epsilon^{bdf}U^{aa'}U^{bb'}U^{cc'}U^{dd'}U^{ee'}U^{ff'}\psi^{a'b'}\psi^{c'd'}\psi^{e'f'} \\
 &= \det U\det U\epsilon^{a'c'e'}\epsilon^{b'd'f'}\psi^{a'b'}\psi^{c'd'}\psi^{e'f'} \\
 &= \epsilon^{a'c'e'}\epsilon^{b'd'f'}\psi^{a'b'}\psi^{c'd'}\psi^{e'f'}
\end{split}
\end{align}
For later convenience we work out the baryon color structure in terms of the fermionic degrees of freedom.
\begin{align}
\begin{split}
 \epsilon^{abc}\epsilon^{a'b'c}\psi^{aa'}\psi^{bb'}\psi^{cc'}
 &= \epsilon^{abc}\epsilon^{a'b'c'}(\tilde{\psi}^ie^i)^{aa'}(\tilde{\psi}^je^j)^{bb'}(\tilde{\psi}^ke^k)^{cc'} \\
 &= [\epsilon^{abc}\epsilon^{a'b'c'}(e^i)^{aa'}(e^j)^{bb'}(e^k)^{cc'}]\tilde{\psi}^i\tilde{\psi}^j\tilde{\psi}^k \\
 &\equiv \xi^{ijk}\tilde{\psi}^i\tilde{\psi}^j\tilde{\psi}^k
\end{split}
\end{align}
Here we define a new contraction symbol $\xi^{ijk}$ with $\{i,j,k\}=1\dots6$. This tensor is symmetric in all indices and it has four independent elements.
\begin{equation}
 \xi^{ijk} = \epsilon^{abc}\epsilon^{a'b'c'}(e^i)^{aa'}(e^j)^{bb'}(e^k)^{cc'}
 \label{eq:xi}
\end{equation}

\section{Numerical results}
\label{sec:appA}
We show in Table \ref{tab:b54} and \ref{tab:b54ratio} the numerical results for the large volume simulations at $\beta=5.4$. \\[1mm]
\begin{table}[H]
\begin{center}
 \resizebox{\linewidth}{!}{%
 \begin{tabular}{c|llllllll}
  $-m_0$ &
  \multicolumn{1}{c}{$L^3\times T$} &
  \multicolumn{1}{c}{$m_{PCAC}$} &
  \multicolumn{1}{c}{$m_{PS}$} &
  \multicolumn{1}{c}{$f_{PS}$} &
  \multicolumn{1}{c}{$m_{V}$} &
  \multicolumn{1}{c}{$f_{V}$} &
  \multicolumn{1}{c}{$m_A$} &
  \multicolumn{1}{c}{$m_N$} \\
  \hline\hline
  1.2750 & $24^3\times40$ & 0.2268(1) & 1.0157(4) & 0.483(8) & 1.095(1) & 0.67(1)  & 1.47(3) & 1.80(1) \\
  1.3000 & $24^3\times40$ & 0.1613(2) & 0.8401(6) & 0.348(12) & 0.905(1) & 0.48(3)  & 1.15(2) & 1.45(1) \\
  1.3125 & $24^3\times40$ & 0.1269(4) & 0.727(2)  & 0.272(13) & 0.781(3) & 0.38(3) & 0.99(2) & 1.26(1) \\
  1.3250 & $24^3\times40$ & 0.0909(2) & 0.586(1)  & 0.200(10) & 0.629(2) & 0.28(1)  & 0.78(2) & 1.00(1) \\
  1.3325 & $24^3\times40$ & 0.0682(2) & 0.481(2)  & 0.150(10) & 0.514(3) & 0.20(2) & 0.66(2) & 0.85(2) \\
  1.3400 & $24^3\times40$ & 0.0436(3) & 0.350(3)  & 0.101(9) & 0.369(6) & 0.14(2) & 0.46(3) & 0.63(2) \\
  1.3425 & $32^3\times48$ & 0.0351(1) & 0.293(1)  & 0.081(5) & 0.308(2) & 0.11(1)  & 0.40(3) & 0.52(2) \\
  1.3450 & $32^3\times48$ & 0.0266(1) & 0.235(1)  & 0.071(4) & 0.249(2) & 0.10(1)  & 0.35(2) & 0.43(2) \\
  \hline
 \end{tabular}}
 \caption{Bare quantities from the large volume simulations at $\beta=5.4$.}
 \label{tab:b54}
\end{center}
\end{table}
\begin{table}[H]
\begin{center}
 \resizebox{\linewidth}{!}{%
 \begin{tabular}{c|llllllll}
  $-m_0$ & $m_{PS}/f_{PS}$ & $m_V/f_{PS}$ & $m_A/f_{PS}$ & $m_N/f_{PS}$  & $f_V/f_{PS}$  & $m_V/m_{PS}$ & $m_A/m_{PS}$ & $m_{PS}L$ \\
  \hline\hline
  1.2750 & 2.10(4)  & 2.27(4)  & 3.01(11) & 3.73(3)  & 1.38(3)  & 1.078(2)  & 1.45(3) & 24.4 \\
  1.3000 & 2.41(7)  & 2.59(9)  & 3.24(13) & 4.16(3)  & 1.39(8)  & 1.077(3)  & 1.36(2) & 20.2 \\
  1.3125 & 2.68(16) & 2.88(17) & 3.64(19) & 4.62(4)  & 1.39(13) & 1.075(6)  & 1.37(2) & 17.6 \\
  1.3250 & 2.94(15) & 3.14(14) & 3.84(21) & 5.00(7)  & 1.41(9)  & 1.073(4)  & 1.34(3) & 14.1 \\
  1.3325 & 3.21(23) & 3.40(24) & 4.30(33) & 5.65(13) & 1.36(15) & 1.068(7)  & 1.37(4) & 11.5 \\
  1.3400 & 3.41(30) & 3.60(34) & 4.32(35) & 6.18(19) & 1.37(18) & 1.056(17) & 1.31(2) & 8.4  \\
  1.3425 & 3.58(19) & 3.75(20) & 4.71(23) & 6.45(22) & 1.34(11) & 1.052(8)  & 1.36(3) & 9.4 \\
  1.3450 & 3.28(18) & 3.44(22) & 4.61(45) & 5.98(26) & 1.39(13) & 1.059(11) & 1.47(6) & 7.5 \\
  \hline
 \end{tabular}}
 \caption{Derived quantities from the large volume simulations at $\beta=5.4$.}
 \label{tab:b54ratio}
\end{center}
\end{table}

\section{Topology}\label{sect:top}
\begin{figure}
\begin{center}
 \includegraphics[width=0.49\textwidth,trim={5mm 0 10mm 0},clip]{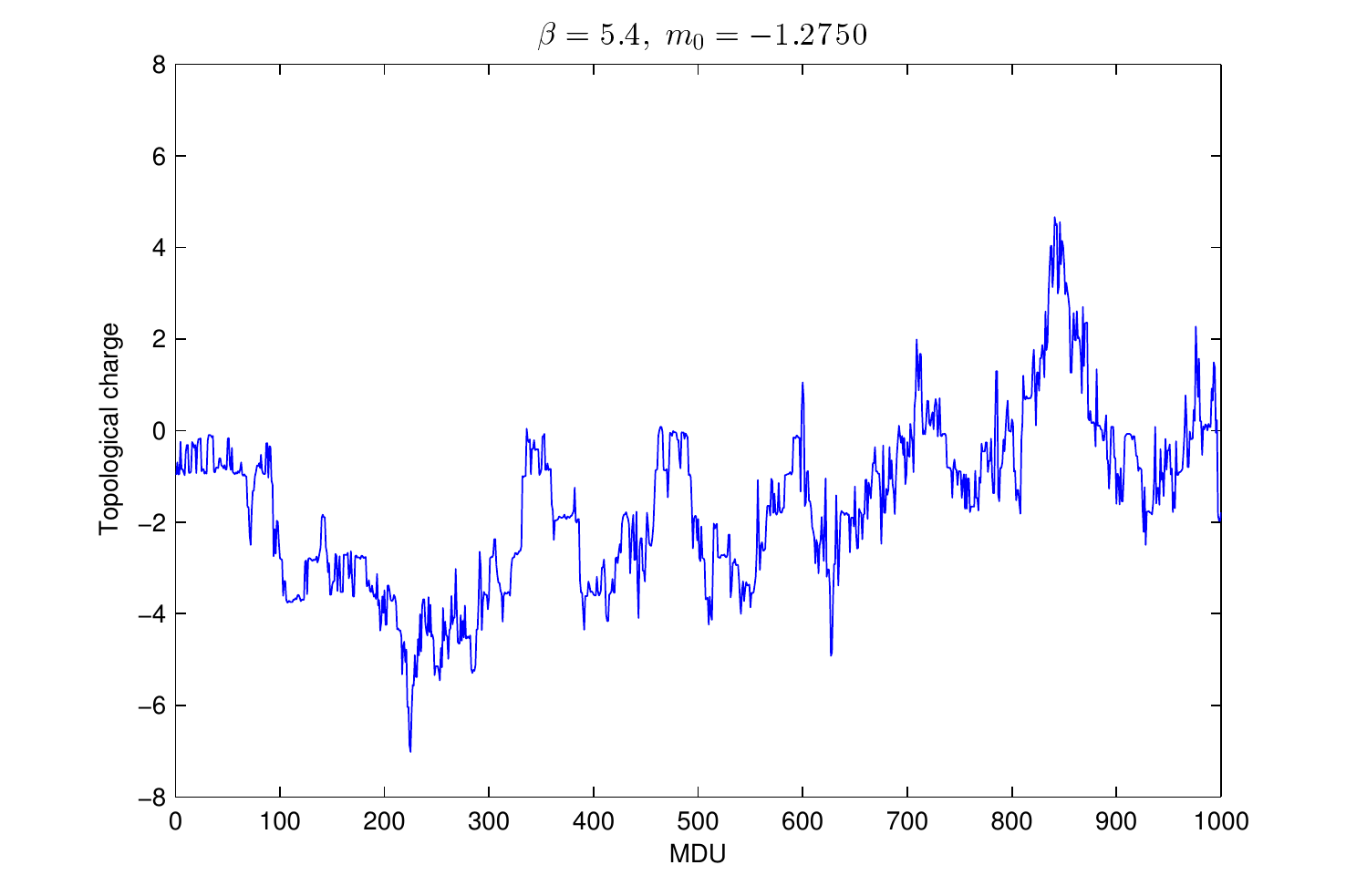}
 \includegraphics[width=0.49\textwidth,trim={5mm 0 10mm 0},clip]{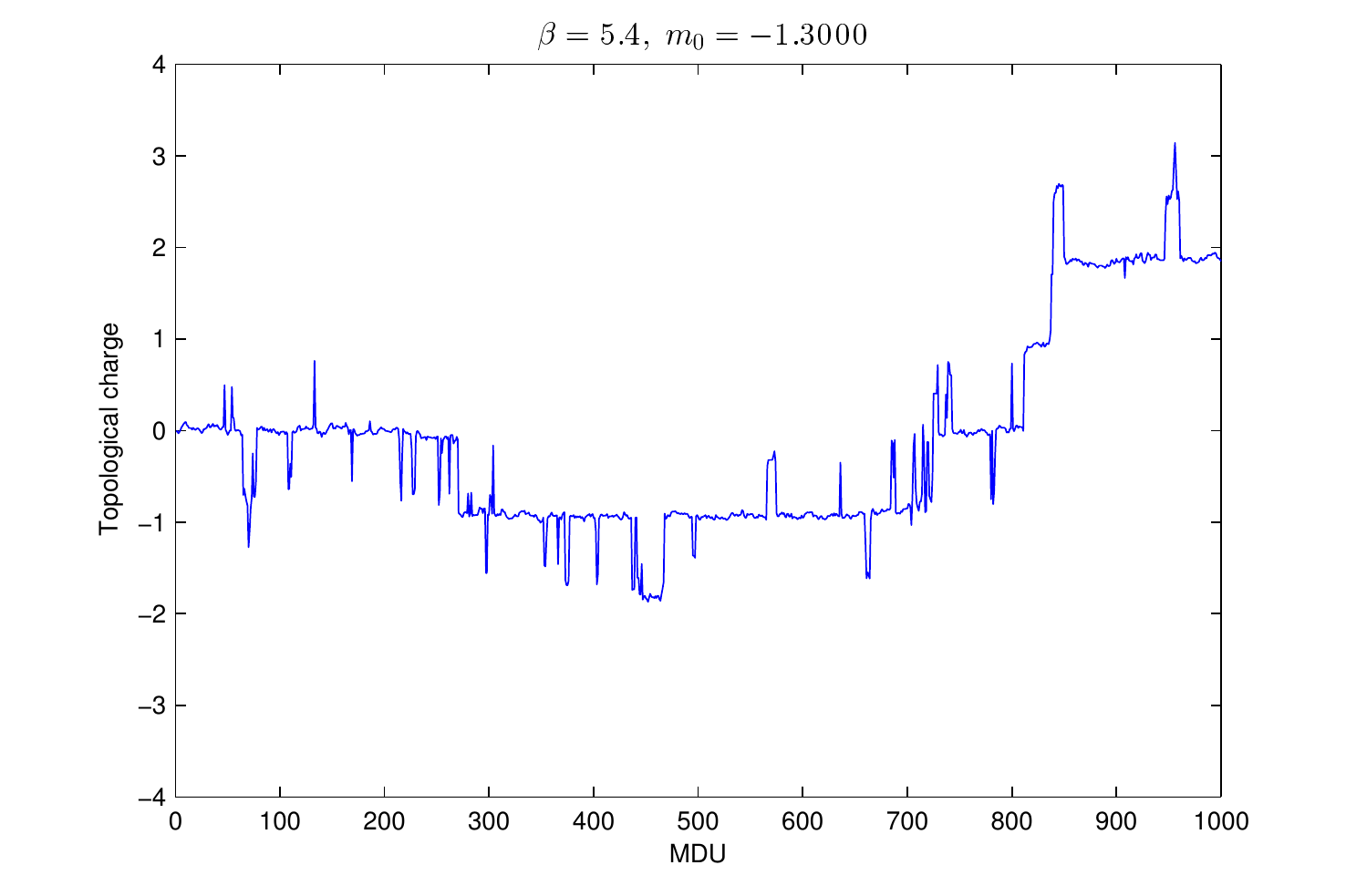} \\
 \includegraphics[width=0.49\textwidth,trim={5mm 0 10mm 0},clip]{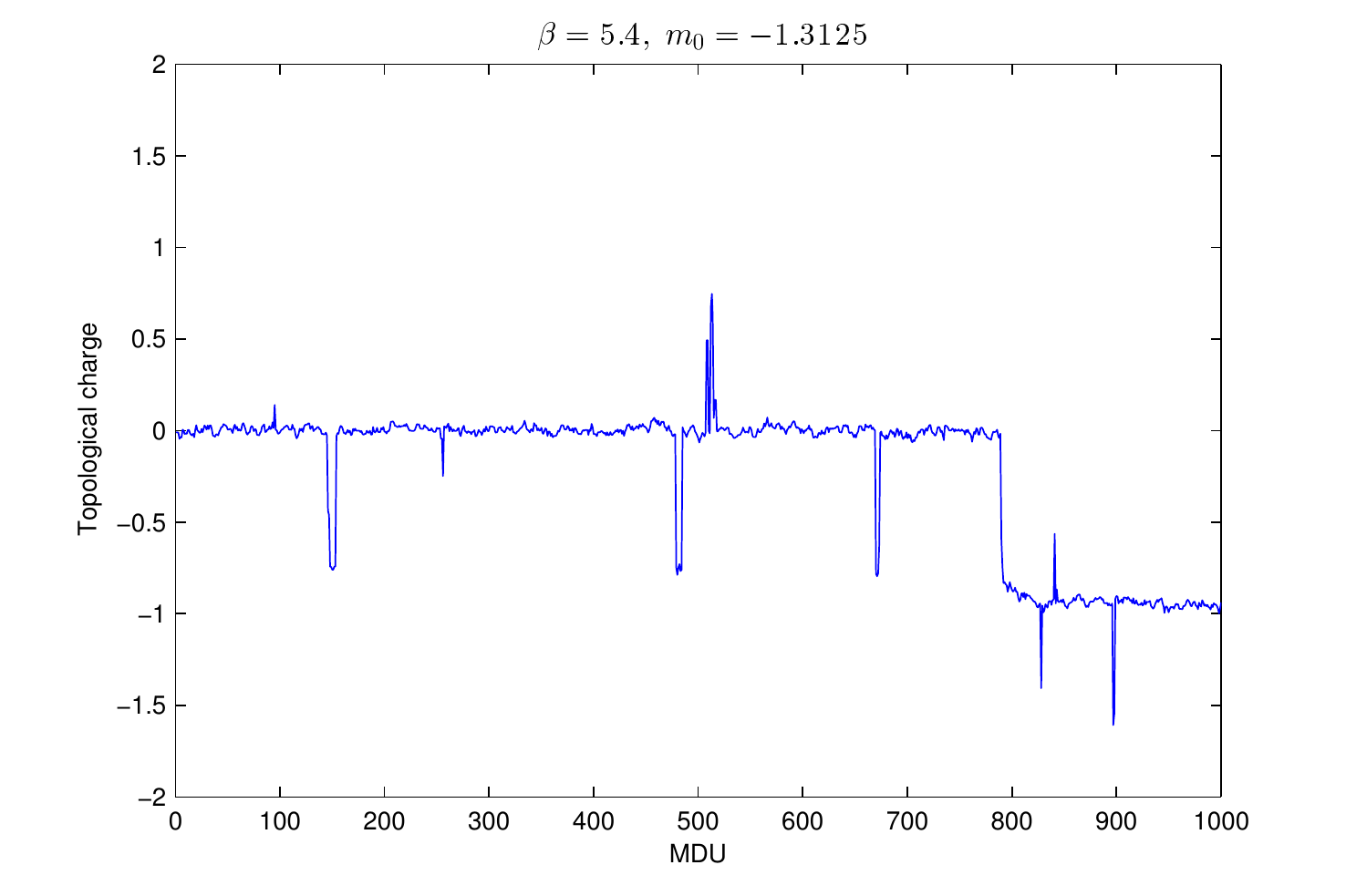}
 \includegraphics[width=0.49\textwidth,trim={5mm 0 10mm 0},clip]{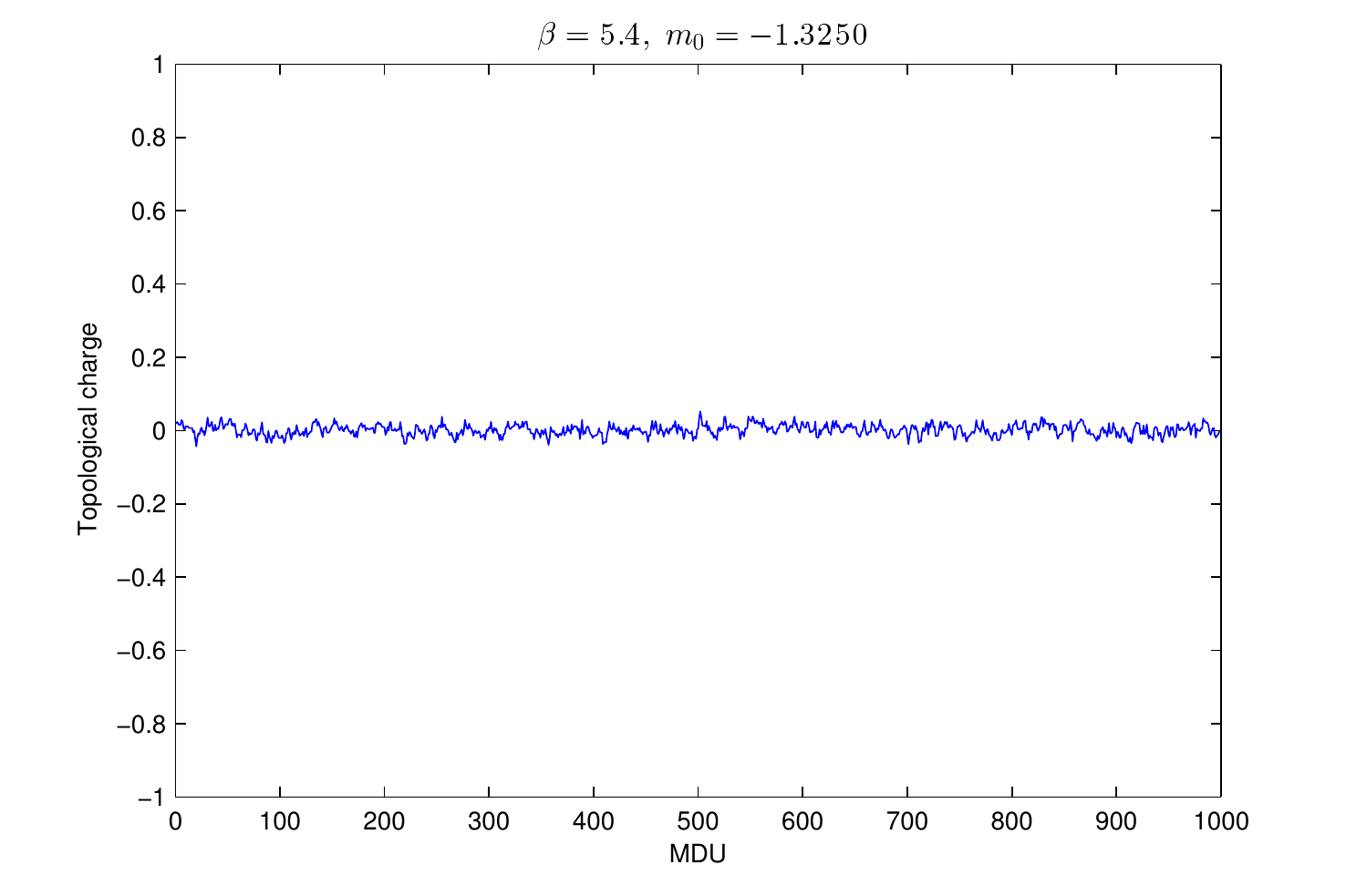} \\
 \vspace{-5mm}
\end{center}
\caption{History of the topological charge for the four heaviest masses in the large volume simulations at $\beta=5.4$. For all lighter masses, the topological charge is always zero.}
\label{fig:topo1}
\end{figure}

As mentioned in the paper, in the weak coupling phase this model suffers from topological freezing to a much larger extent than QCD. In Fig.~\ref{fig:topo1} we show the history of the topological charge for the four heaviest masses in the large volume simulations at $\beta=5.4$. For sufficiently heavy masses we do observe some fluctuation in the topological charge, but as we approach the chiral limit, the topological charge freezes completely.

\begin{figure}
\begin{center}
 \includegraphics[width=0.49\textwidth,trim={5mm 0 10mm 0},clip]{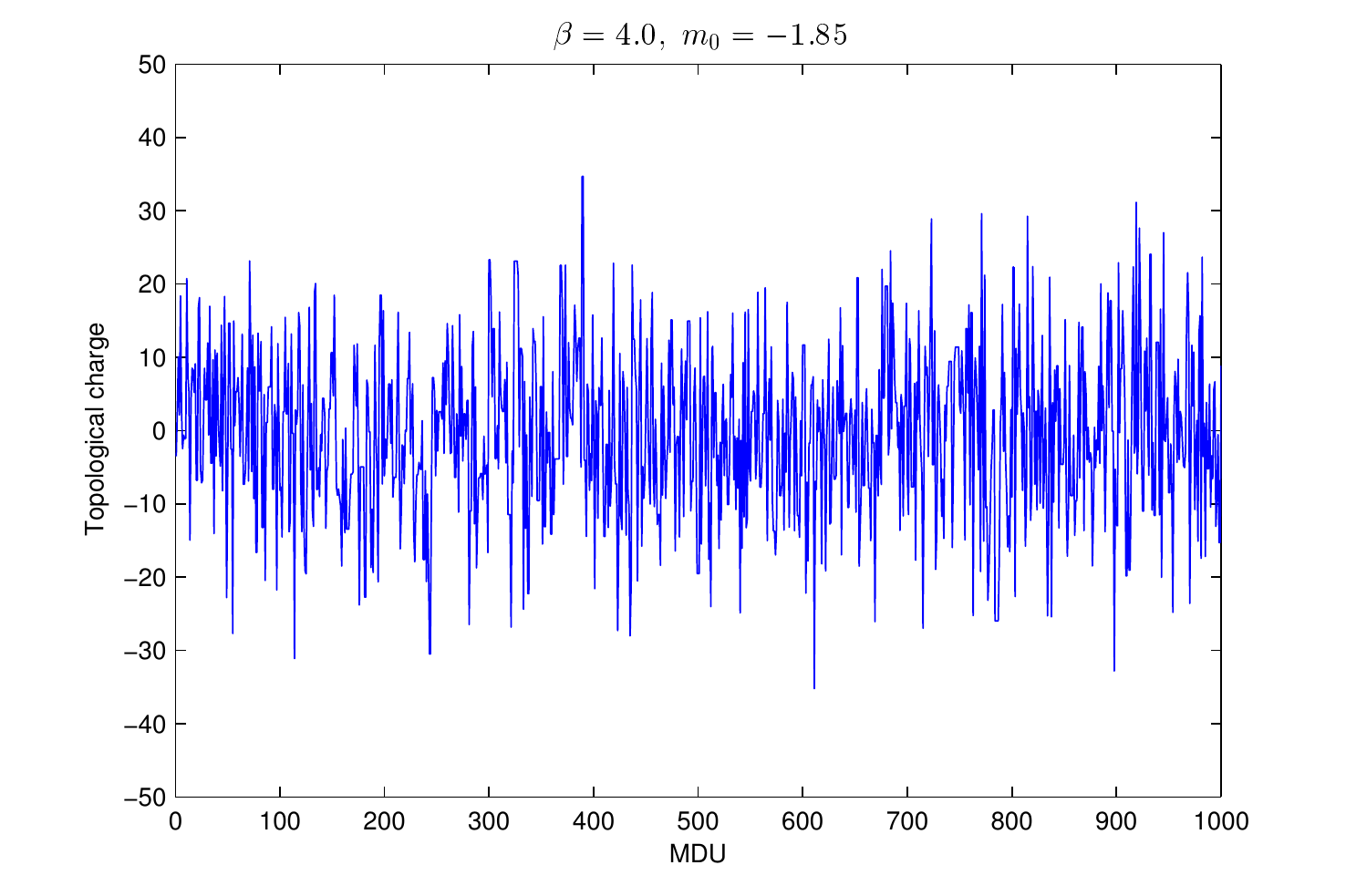}
 \includegraphics[width=0.49\textwidth,trim={5mm 0 10mm 0},clip]{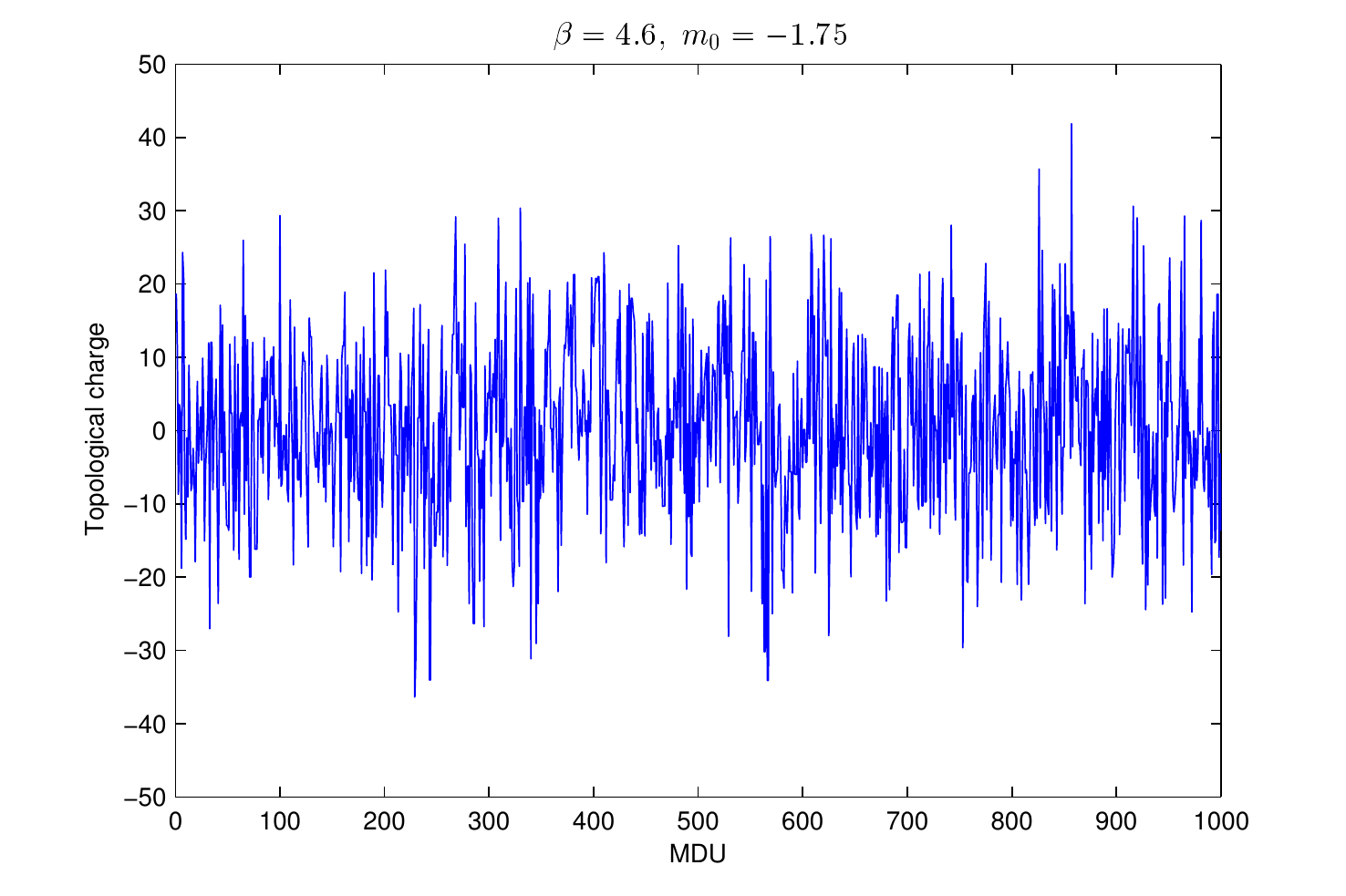} \\
 \includegraphics[width=0.49\textwidth,trim={5mm 0 10mm 0},clip]{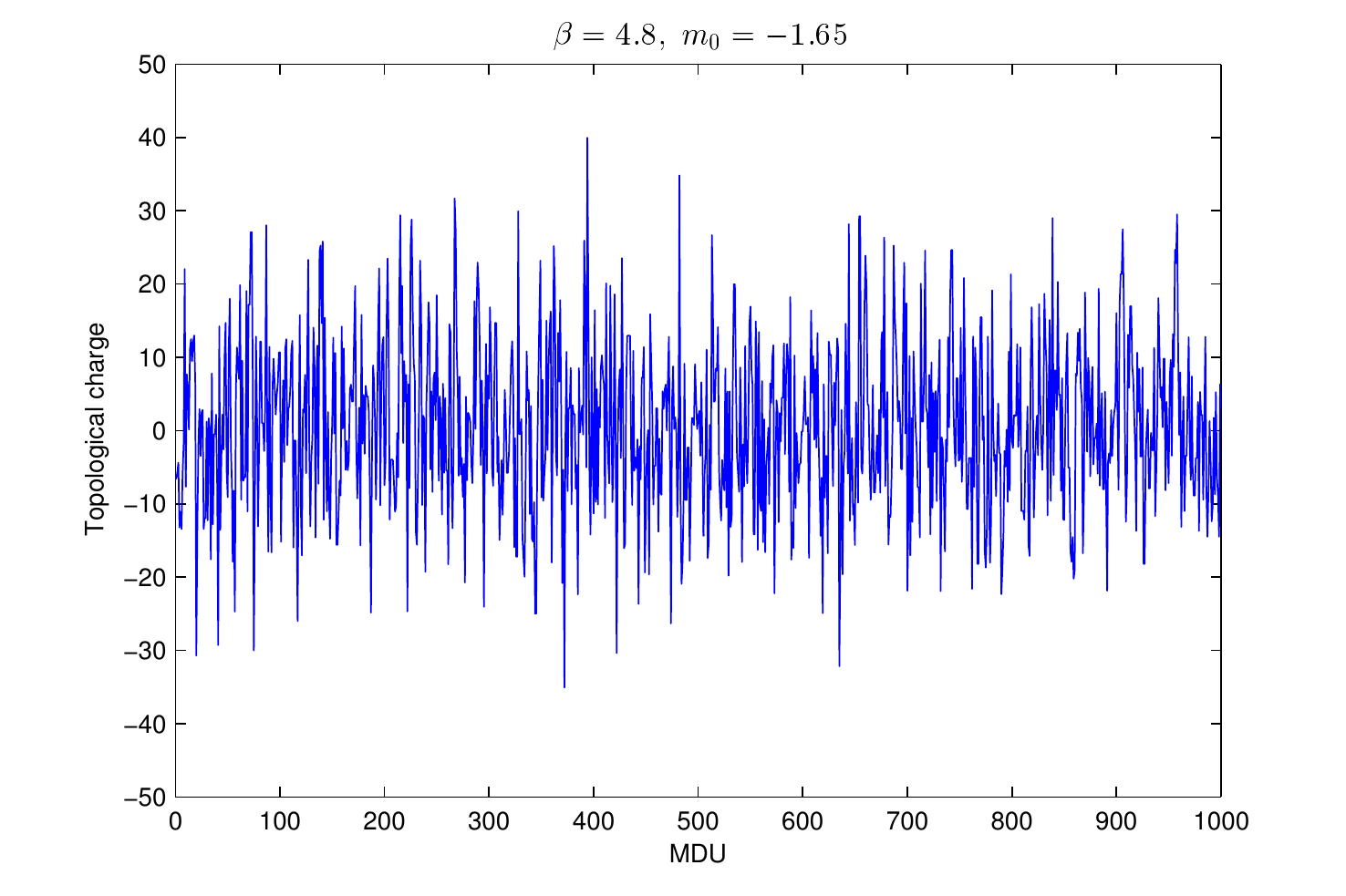}
 \vspace{-5mm}
\end{center}
\caption{History of the topological charge for different simulations in the strong coupling region. The plots show the result for the lightest available mass for three different values of the bare coupling.}
\label{fig:topo2}
\end{figure}

In the strong coupling phase, this does not appear to be a problem. In Fig.~\ref{fig:topo2} we show the history of the topological charge for the lightest available mass at three different bare couplings. This is again an indication that something drastic happens when moving from the strong to the weak coupling phase.

\end{document}